\documentclass[final]{dmtcs-episciences}

\usepackage[round]{natbib}
\usepackage{tikz,pgfplots}
\usetikzlibrary{shapes.geometric}
\usepackage{wrapfig}
\usepackage[utf8]{inputenc}
\usepackage{csquotes}
\usepackage{thmtools}
\usepackage{thm-restate}
\usepackage{url}
\usepackage{hyperref}

\usepackage{subfigure}

\usepackage{amssymb,amsmath,amsfonts}
\usepackage{amsthm}

\title{Three Hardness Results for Graph Similarity Problems
}

\author{
 He Sun\affiliationmark{1,2} \and Danny Vagnozzi\affiliationmark{3}}
\affiliation{
Shenzen Institue of Advanced Technology, Chinese Academy of Sciences, Shenzen, China\\
University of Edinburgh, Edinburgh, United Kingdom \\
Durham University, Durham, United Kingdom
}

\publicationdata{vol. 28:2}{2026}{36}{10.46298/dmtcs.12508}{2023-11-03; 2023-11-03; 2026-05-08}{2026-06-12}

\newcommand{\Inj}{\mathrm{Inj}}
\newcommand{\NP}{\textsf{NP}}
\newcommand{\MMC}{\mathtt{MMC}}
\newcommand{\DIST}{\mathtt{DIST}}

\newcommand{\edit}{\mathcal{E}}
\newcommand{\Res}{\mathrm{Res}}
\newcommand{\C}{\mathcal{C}}

\newcommand{\sgn}{\mathsf{sgn}}

\newcommand{\even}{\mathtt{even}}
\newcommand{\odd}{\mathtt{odd}}

\definecolor{newgreen}{rgb}{0.0, 0.42, 0.24}

\newtheorem{thm}{Theorem}[section] 

\newtheorem{lem}[thm]{Lemma}

\newtheorem{defi}[thm]{Definition}

\newtheorem{cor}[thm]{Corollary}
\newtheorem{fac}[thm]{Fact}

\newtheorem{pro}[thm]{Proposition}

\newtheorem{openproblem}[thm]{Open Problem}
\begin{document}

\maketitle

\begin{abstract}
 
Notions of graph similarity   provide alternative perspective on the graph isomorphism problem and vice-versa. In this paper, we consider measures of similarity arising from mismatch norms as studied by 
Gervens and Grohe: the edit distance \begin{math}\delta_\edit\end{math}, and the metrics  arising from \begin{math}\ell_p\end{math}-operator norms, which we denote by \begin{math}\delta_p\end{math} and \begin{math}\delta_{|p|}\end{math}. We address the following question: can these measures of similarity be used to design polynomial-time approximation algorithms for graph isomorphism? We show that computing the value of  \begin{math}\delta_\edit\end{math} is \NP-hard on pairs of graphs with the same number of edges.  In addition, we show that computing \begin{math}\delta_p\end{math} and \begin{math}\delta_{|p|}\end{math} is \NP-hard even on pairs of \begin{math}1\end{math}-planar graphs with the same degree sequence and bounded degree. These results improve on previous known ones. Indeed the complexity of computing \begin{math}\delta_\edit\end{math} has not been considered for pairs of graphs with the same number edges. Furthermore, whilst the hardness proof of computing \begin{math}\delta_p\end{math} and \begin{math}\delta_{\left|p\right|}\end{math} provided by Gervens and Grohe extends to pairs of graphs with same number of edges, we strengthen this result by showing hardness on pairs of graph with same degree sequence. Our proof involves some technical analysis of the spectra of adjacency matrices of signed graphs, which may be of interest independently. Finally, we study similarity problems on strongly regular graphs and prove some quasi-optimal inequalities regarding our measures of similarities, with interesting consequences on the computational complexity of group isomorphism.
\end{abstract}

\section{Introduction} 
 The \emph{graph isomorphism problem} is a classical problem in theoretical computer science. It consists in determining whether there is an edge preserving bijection between the vertices of two input graphs. While there are several applications of graph isomorphism algorithms in  real-world settings such as  pattern recognition and computer vision, most studies on the graph isomorphism problem have been motivated by purely theoretical intricacies. On the one hand, while we do not know whether the problem is in \textsf{P}, it is believed not to be   \NP-complete. On the other hand, following a breakthrough by Babai~\citeyearpar{babai:babai}, we know that graph isomorphism is decidable in quasipolynomial time. Adding this to the fact that many practical instances are solvable in polynomial time,  one can safely claim that on average the  graph isomorphism problem can be solved efficiently.

In comparison to graph isomorphism, \emph{graph similarity} is a more general concept and has received 
   plentiful attention in practical applications~\citep{pr1,pr2,pr3}; however its  studies from a theoretical point of view have remained limited.  Informally speaking, graph similarity is a family of decision problems, and every such problem is associated with some measure of similarity, for example a \emph{graph metric} $\delta$ mapping pairs of graphs into non-negative real numbers. For a fixed metric \begin{math}\delta\end{math}, the input of the problem consists in a pair of graphs \begin{math}G,H\end{math} and a real constant \begin{math}c\end{math}, and it is asked to decide whether \begin{math}\delta(G,H)\leq c\end{math} holds. As an example, consider a metric $\iota$ assigning $0$ to pairs of isomorphic graphs and $1$ otherwise. Evaluating $\iota$ is equivalent to deciding graph isomorphism, which is therefore in itself a similarity problem. Another interesting metric that has appeared in the literature under different guises is the \emph{edit distance} \citep{grohe2018}, which we denote by $\delta_\edit$. Given two graphs $G$ and $H$ with the same number of vertices, $\delta_\edit(G,H)$ is defined to be the minimum number of edges to be deleted from and added to $H$ in order to obtain a graph that is isomorphic to $G$. We call $\DIST_\edit$ the similarity problem associated to the metric $\delta_\edit$. This problem is known to be \NP-hard even  on some very restricted classes of graphs, and is closely related to several   computationally hard problems such as the maximum common edge subgraph problem or the quadratic assignment problem, both of which are well-known to be notoriously hard optimisation problems.

Very recent work by Gervens et al.~\citep{grohe_similarity} studies the computational complexity of graph similarity problems,  and builds a theoretical framework for studying such problems. This includes some precise definitions of graph metrics as well as analysing the metrics from matrix norms that are  commonly used in graph theory.
These metrics  are all based on minimising some measure $\mu$ of `mismatch' between two graphs, which can be seen as the discrepancy between the graphs for a given alignment. The edit distance can be classed as such a metric. Indeed, for a bijection $\pi:V(G)\rightarrow V(H)$, let $G^\pi$ be the image of $G$ under $\pi$,  and the define the \emph{edit norm} $\mu_\edit(G^\pi,H)$   to be the number of edges $(u,v)$ of $G$ for which $\pi$ is not a local isomorphism; that is, $\left(\pi(u),\pi(v)\right)$ is an edge if, and only if, $(u,v)$ is not an edge. Then the edit distance $\delta_\edit(G,H)$ is given by the minimum value of $\mu_\edit(G^\pi,H)$ over all bijections $\pi$. In a more abstract sense, this can be seen as minimising the number of edges in the \emph{mismatch graph} $G^\pi-H$, a signed graph whose adjacency matrix is given by the difference between the adjacency matrix of $G^\pi$ and that of $H$. The other graph metrics analysed in \citep{grohe_similarity} also consist in minimising some property of a mismatch graph. In particular, they consider the (signed and unsigned) $\ell_p$-\emph{operator norm} and the \emph{Lovasz cut norm} of the mismatch graph as a measure for mismatch. In this paper we are only concerned with the edit norm and the (signed and unsigned) $\ell_p$-operator norms, which are natural generalisations of the Euclidean norm and will be formally defined in Section \ref{sec:pre}. The main difference between these metrics is that the value of the latter can be estimated by the maximum degree of the mismatch graph, hence a \emph{local} property, whereas the former seems to rely mostly on \emph{global} properties of the mismatch graph. Nonetheless, these metrics give rise to \NP-hard similarity problems, and the strategies to prove the hardness result are strikingly similar; that is, by reducing the \NP-hard Hamiltonian cycle problem on $3$-regular graphs to the similarity problem between a $3$-regular graph and a cycle of the same order. The hardness proof of the similarity problem associated with the Lovasz cut norm follows a different strategy \citep{grohe_similarity} and will not be considered in this paper.

A natural question arising from these proofs is whether \NP-hardness still holds when considering the similarity problem on pairs of graphs harder to distinguish isomorphism-wise than, say, a $3$-regular graph and a cycle. In fact, the main question motivating this paper is the following: \emph{are there classes of graphs where some notion of similarity can be used as an efficient approximation of isomorphism?}
Put otherwise, we ask whether there are classes of graphs whose isomorphism is hard to decide but one can decide in polynomial time whether $\delta(G,H)\leq t$ for some metric $\delta$ and threshold $t$ (which could be constant or, for example, a function of the number of vertices), so we can deduce that $G$ and $H$ are not isomorphic if $\delta(G,H)>t$, and deem the test inconclusive otherwise. Since 3-regular graphs are distinguished from cycles using simple properties such as the number of edges or their degree sequences, the known hardness proofs for the metrics studied in \citep{grohe_similarity} do not answer the above question.

We address this issue in light of the graph metrics studied in \citep{grohe_similarity}. Let $\delta_\edit,\delta_p$ and $\delta_{|p|}$ be the graph metrics arising from the edit distance and the $\ell_p$-operator norms of the signed and unsigned mismatch graph respectively. Denote the similarity problem associated with each of these metrics by $\DIST_\edit, \DIST_p$ and $\DIST_{|p|}$. The following are our main results.
 \begin{thm}\label{thm:main_2}
 For any constant $c>0$ and $ \epsilon\in(0,1/2)$, the following holds: for any pair of graphs $G$ and $H$ with $n$ vertices and the same number of edges, it is \NP-hard to decide whether $\delta_\edit(G,H)\leq cn^\epsilon$. Consequently, $\DIST_\edit$ is \NP-hard, even when the two input graphs have the same number of edges.
   % $\DIST_\edit$ is \NP-hard even when restricted to pairs of graphs with equal volumes. In particular, for any $0<\epsilon< 1/2$ and any $t(n)=\Theta\left(n^\epsilon\right)$, it is \NP-hard to decide whether $\delta_\edit(G,H)\leq t(n)$ for pairs of graphs $G,H$ with equal volumes.
\end{thm}

\begin{thm}\label{thm:main_1}
    For any pair of $1$-planar graphs $G$ and $H$ with $n$ vertices, the same degree sequence and maximum degree at most $15$, it is \NP-hard to decide whether $\delta_p(G,H)\leq 2$ and $\delta_{|p|}(G,H)\leq 2$ for any $p\geq 1$. Consequently, $\DIST_p$ and $\DIST_{|p|}$ are \NP-hard, even when the two input graphs have bounded degree and the same degree sequences.
\end{thm}

The construction employed in the proof of Theorem \ref{thm:main_1} is interesting in a twofold manner.  Firstly, we remark that a priori, it is not clear if these graphs can be used to show the hardness of $\DIST_\edit$ for pairs of graphs with same degree sequence. This is in contrast to the more general problem, where considering as input a $3$-regular graph and a cycle graph is sufficient to show the hardness of $\DIST_\edit$, $\DIST_p$ and $\DIST_{|p|}$. Secondly, the proof for $p=2$ differs significantly from the proof for values of $p\neq 2$, and requires a deep analysis of the combinatorial properties of signed graphs. These anomalies seem to point at an interesting layer of discrepancy between these similarity problems.

While the statement of Theorem \ref{thm:main_2} seems to diminish the hopes of a threshold function $t$ asymptotically smaller than $\sqrt{n}$ such that $\delta_\edit(G,H)\leq t(n)$ is an efficiently decidable approximation of isomorphism, neither Theorem~\ref{thm:main_2} nor Theorem \ref{thm:main_1} provide a satisfactory answer to our motivating question. In fact, the pairs of graphs used in their proofs are still distinguishable by simple properties such as their 
degree sequences~(Theorem~\ref{thm:main_2}) or 
spectra~(Theorem~\ref{thm:main_1}). As an attempt to study the complexity of $\DIST_p$ and $\DIST_{|p|}$ over pairs of cospectral graphs (that is, whose adjacency matrices have the same eigenvalue spectrum), we prove another hardness statement with a similar effect but in a different style.
\begin{thm}\label{thm:main_3}
Let $G$ and $H$ be cospectral graphs on $n$ vertices each with maximum degree $d_{\max}$, and consider some function $t(n)=o\left(n^{1/2}\right)$. The following hold:
\begin{enumerate}
    \item If there is a polynomial-time algorithm deciding whether $\delta_\edit(G,H)\leq t(n)$, then there is a polynomial-time algorithm deciding the isomorphism of groups as Cayley tables \footnote{When it comes to group isomorphism, it is important to specify the format in which the groups are given as an input. For example, if the groups are given as a list of generators and relations thereof, then group isomorphism is undecidable~\citep{dehn}.}.
    \item If there is a polynomial-time algorithm deciding whether $\delta_1(G,H)\leq d_{\max}/3-4$, then there is a polynomial-time algorithm deciding the isomorphism of groups as Cayley tables \footnote{This statement is essentially the same as the previous one. However, the bound is written in terms of $d_{\max}$ rather than $n$, since $\delta_1$ is has close ties to $d_{\max}$. More precisely, for any two graphs $G$ and $H$ with maximum degree $d_{\max}$, it trivially follows from the definition of $\delta_1$ (see Section 2.1) that $\delta_1\left(G,H\right)\leq 2d_{\max}$.}.
\end{enumerate}
\end{thm}

This follows from a simple combinatorial analysis for bounds on alignments between strongly regular graphs (Proposition \ref{prop:srg}). Whilst this analysis might seem na\"ive at a first glance, we show in Appendix~A that our obtained bound is almost tight up to some constant additive term. 

Finally, we  comment on Theorem \ref{thm:main_3} in light of current knowledge on group isomorphism. The fastest known algorithm for group isomorphism runs in quasipolynomial time. However, it is believed that an efficient algorithm for group isomorphism could help overcome some of the current bottlenecks in Babai's quasipolynomial time algorithm for graph isomorphism~\citep{babai:babai}. As such, Theorem~\ref{thm:main_3} suggests that the existence of a small threshold function $t$ so that $\delta_\edit(G,H)\leq t(n)$, or a constant $c$ such that $\delta_1(G,H)\leq c$ provide efficiently decidable approximations of isomorphism is unlikely.

\subsection{Related Work}

Our results have close ties to those in \citep{arvind,grohe_similarity,grohe2018}. The authors in~\citep{arvind} study the $\DIST_\edit$ problem in the context of approximating graph isomorphism. The authors provide both hardness results and efficient algorithms for different variants of the problem of approximating some optimisation version of the $\DIST_\edit$ problem. In \citep{grohe_similarity,grohe2018}, the authors show the hardness of the problems $\DIST_\edit$, $\DIST_p$ and $\DIST_{|p|}$ for all natural numbers $p$. In addition, for $\DIST_p$ and $\DIST_{\left|p\right|}$ hardness is obtained even when restricted to forests of bounded degree. However, the classes of graphs resulting from these reductions include pairs of graphs with differing degree sequences. Thus, whilst the proofs of Theorems \ref{thm:main_2} and \ref{thm:main_1} are inspired from the results in \citep{grohe_similarity,grohe2018}, they  provide a different perspective on the nature and computational complexity of graph similarity problems.

Linear algebraic approaches to similarity prior to the $\ell_p$-operator norms in \citep{grohe_similarity} have appeared in \citep{kolla, umeyama}. For instance, in~\citep{kolla} the authors consider the \emph{spectrally robust graph isomorphism problem} - a similarity problem based on the eigenvalues of Laplacian matrices.

Related to similarity problems on Latin square graphs are the works on the Hamming distance between multiplication tables of finite groups~\citep{denes,praeger, drapal}. More recent work~\citep{buchheim} provides hardness results for variants of the \emph{subgroup distance problem}.

\subsection{Organisation}

The remaining part of the paper is organised as follows: we list the notation and background knowledge in Section~\ref{sec:pre}. Section~\ref{sec:volume} studies the computational complexity of the $\DIST_\edit$ problem, and proves Theorem~\ref{thm:main_2}; Section~\ref{sec:2ndmain} studies the computational complexity of the  $\DIST_p$ and $\DIST_{|p|}$ problems, and proves Theorem~\ref{thm:main_1}. We study the  similarity problem for strongly regular graphs, and prove Theorem~\ref{thm:main_3} in Section~\ref{sec:srg}. Section~\ref{sec:open} provides some natural open questions to finish the discussion. %All the omitted technical details can be found in the appendix.

\section{Preliminaries\label{sec:pre}}

\subsection{Notation}
 
For a graph $G$ on $n$ vertices, we denote its vertex set by $V(G)$ and its edge set by $E(G)$. We only consider simple undirected graphs, so the edges $(u,v)$ and $(v,u)$ are identical. The \emph{adjacency matrix} of $G$ is denoted by $A_G\in\mathbb{R}^{V\times V}$, where $A_{uv}=1$  if $(u,v)\in E(G)$, and $A_{uv}=0$ otherwise. Given a vertex $v$, we indicate its \emph{neighbourhood} in $G$ by $N_G(v)$; that is, $N_G(v)=\left\{u\in V\mid (u,v)\in E(G)\right\}$.

For a mapping $\pi:S\rightarrow T$ and a subset $S'\subseteq S$, denote by $\pi\left(S'\right)$ the image $\pi$ restricted to $S'$; that is, $\pi\left(S'\right)=\left\{\pi(s)\mid s\in S'\right\}$. Given any matrix $A\in\mathbb{R}^{V\times V }$ and an injective mapping $\pi:V\rightarrow W$, we denote by $A^\pi$ the $\pi(V)\times \pi(V)$ matrix with entries $A_{\pi(v)\pi(w)}=A_{vw}$.  If $V=V(G)$ for some graph $G$, we define $G^\pi$ to be the graph whose adjacency matrix is $A_G^\pi$. That is, $V(G^\pi)=\pi\left(V(G)\right)$ and $E(G)=\left\{(\pi(u),\pi(v))\mid (u,v)\in E(G)\right\}$. Following the notation and terminology in \citep{grohe_similarity}, we denote by $\Inj(G,H)$ the set of injective maps $\pi:V(G)\rightarrow V(H)$ for graphs $G$ and $H$. If $\pi$ is bijective, we refer to it as an \emph{alignment}. In particular, $G$ and $H$ are isomorphic if and only if there is an alignment $\pi$ for which $G^\pi=H$.

\begin{figure}[ht]
    \centering
   \resizebox{8cm}{!}{  
 \begin{tikzpicture}[main/.style = {draw, circle, inner sep=0pt, minimum size=10pt}][scale=2]
    \node[draw,  circle,color=newgreen, fill=green!5] (u) at (-2,0) {$u_1$};
    \node[draw, circle, color=newgreen, fill=green!5] (v) at (0,0) {$u_2$};
    \node[draw, circle, color=newgreen, fill=green!5] (w) at (2,0) {$u_3$};
    \node[draw, circle, color=newgreen, fill=green!5] (x) at (4,0) {$u_4$};
    \draw[color=black,  very thick] (u)--(v)--(w);

    \node[draw, circle, color=newgreen, fill=green!5 ] (a) at (-2,-1.5) {$v_1$};
    \node[draw, circle, color=newgreen, fill=green!5] (b) at (0,-1.5) {$v_2$};
    \node[draw, circle, color=newgreen, fill=green!5] (c) at (2,-1.5) {$v_3$};
    \node[draw, circle, color=newgreen, fill=green!5] (d) at (4,-1.5) {$v_4$};
    \draw[color=black, very thick] (d)--(c)--(b);

    \node[draw, circle, color=newgreen, fill=green!5] (a') at (-2,-3) {$v_1$};
    \node[draw, circle, color=newgreen, fill=green!5] (b') at (0,-3) {$v_2$};
    \node[draw, circle, color=newgreen, fill=green!5] (c') at (2,-3) {$v_3$};
    \node[draw, circle, color=newgreen, fill=green!5] (d') at (4,-3) {$v_4$};
    \draw[color=blue, very thick] (a')--(b');
    \draw[color=red, very thick] (c')--(d');

    \node (G) at (-4,0) {$G$};
    \node (H) at (-4,-1.5) {$H$};
    \node (G-H) at (-4,-3) {$G^\pi-H$};
    \end{tikzpicture}}
    \caption{Construction of $G^\pi-H$ based on $G$, $H$, and  $\pi$ mapping $u_i\mapsto v_i$. Here, the edges $\left(v_1,v_2\right)$ and $\left(v_3,v_4\right)$ are positive and negative for $\pi$ respectively.}
    \label{fig:gpih}
\end{figure}
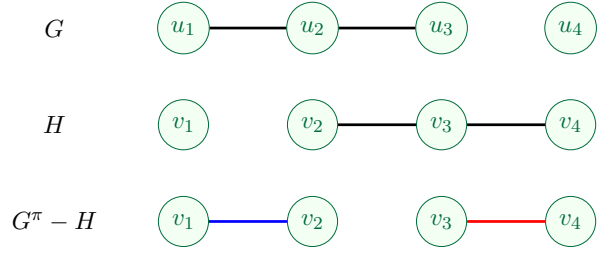

Let $G$ and $H$ be graphs with the same vertex set $V$. The graph $G-H$ is defined to have vertex set $V$ and and edge set $E(G)\backslash E(H)\cup E(H)\backslash E(G)$, where the edges in $E(G)\backslash E(H)$ have weight $+1$ and the ones in $E(H)\backslash E(G)$ have weight $-1$. For more general graphs $G$ and $H$ with the same number of vertices and some alignment $\pi\in \Inj(G,H)$, we call $G^\pi- H$ the \emph{mismatch graph} of $\pi$, a concept first introduced in \citep{grohe_similarity}.  We refer to the elements of $E(G^\pi)\backslash E(H)$ as being \emph{positive} for $\pi$, and the ones of $E(H)\backslash E(G^\pi)$ as being \emph{negative} for $\pi$; see Figure~\ref{fig:gpih} for an illustration of graph $G^{\pi}-H$ based on $G,H$ and $\pi$.

\vspace{0.2cm}

A graph whose edges have weight $\pm1$ is sometimes referred to as a \emph{signed graph}. Intuitively, the negative and positive edges are the ones to be removed from and added  to $H$ respectively so as to obtain the graph $G^\pi$. In the spirit of the above, we refer to the elements of $E(G^\pi)\cap E(H)$ as being \emph{neutral} for $\pi$. Note that $V(H)$ together with the neutral and positive edges of $G^\pi - H$ form the graph $G^\pi$, and $V(H)$ together with the neutral and negative edges of $G^\pi - H$ form the graph $H$. Furthermore, one may also note that an edge $(u,v)$ is positive (or negative) if and only if the $(u,v)$ entry of $A_{G^\pi}-A_H$ is $+1$ (or $-1$ respectively). In particular, the adjacency matrix of the mismatch graph $G^\pi- H$ is $A_{G^\pi}-A_H$. Since $G^\pi-H$ and $H$ have the same vertex set by definition, we are careful to indicate which graph we are referring to when treating some vertex $v\in V(H)$. While $G^\pi-H$ is effectively a weighted graph, the \emph{degree} of a vertex $v$ in $G^\pi-H$ is to be understood as the number of edges incident on $v$ irrespective of the sign; moreover, the  \emph{volume} of $G^\pi-H$ is defined as  the sum of the degrees of all vertices, so twice the number of edges in the graph. The following fact is easy to verify, and is used throughout the paper.

\begin{fac}\label{fact:pos_neg}
If $G$ and $H$ are both $d$-regular graphs with the same number of vertices, then for any $\pi\in\Inj(G,H)$ the vertices in $G^\pi-H$ have even degree. In particular, the number of positive edges incident on a vertex is equal to the number of negative edges. 
\end{fac}

\subsection{Graph Metrics Based on Mismatch Norms}

To formally study graph metrics, the authors of~\citep{grohe_similarity} introduce the notion of a \emph{mismatch norm $\mu$}. For a pair of graphs $G$ and $H$ with the same vertex set, $\mu(G,H)$ is defined as some function of $G-H$. Notice that two isomorphic copies of $G$ do not necessarily give rise to the same signed graph, so the metric $\delta(G,H)$ is defined to be the minimum value of the mismatch norm over all isomorphic copies of $G$. In this work we examine the following mismatch norms:
%In \cite{grohe_similarity}, the authors introduced the concept of a \emph{mismatch norm} $\mu$ between pairs of graphs, giving rise to some idea of distance, or more precisely, a \emph{graph metric} $\delta$. Intuitively, for a pair of graphs $G$,$H$ with the same vertex set, $\mu(G,H)$ measures some notion of `distance' as a function of $G-H$. Note that two isomorphic copies of $G$ do not give rise to the same signed graph $G-H$, so the corresponding metric $\delta(G,H)$ is defined to be the minimum value of the mismatch norm over all isomorphic copies of $G$. In this paper, we are concerned with the following mismatch norms (in the definitions that follow, assume for simplicity that $G$ and $H$ have the same vertex set):
\begin{enumerate}
    \item The \emph{edit norm}: \[\mu_\edit(G,H):=\left|E(G)\backslash E(H)\cup E(H)\backslash E(G)\right|.\]
    Note that this norm counts the number of edges in $G-H$.
    \item The $\ell_p$-\emph{operator norm}: \[\mu_p(G,H):=\sup_{x\in \mathbb{R}^n\backslash\{0\}}\frac{|\!|(A_G-A_H)x|\!|_p}{|\!|x|\!|_p},\] where $|\!|x|\!|_p$     
    for a vector $x\in \mathbb{R}^V$ is defined as $|\!|x|\!|_p:=\left(\sum_{v\in V}|x_v|^p\right)^{1/p}$. For $p=2$, this norm is sometimes referred to as the \emph{spectral norm}, and coincides with the maximum absolute value of the eigenvalues of the matrix in the argument.
    \item The \emph{absolute $\ell_p$-operator norm}: \[\mu_{|p|}(G,H):= |\!|\mathbf{abs}(A_G-A_H)|\!|_p, \]where the operator $\mathbf{abs}$ takes componentwise absolute values. As above, for $p=2$ this coincides with the supremum of the absolute values of the eigenvalues of $\mathbf{abs}\left(A_G-A_H\right)$.
\end{enumerate}

We define $\delta_\edit, \delta_p,\delta_{|p|}$ to be the graph metrics corresponding to the above three mismatch norms, respectively. That is, \[
\delta_\edit(G,H):=\min_{\pi\in \Inj(G,H)}\mu_\edit(G^\pi,H),
\]and we define $\delta_p,\delta_{|p|}$ in the same way.  Let $\DIST_\edit$  be the problem of deciding whether $\delta_\edit(G,H)<c$ for some parameter $c$, and similarly define  $\DIST_{|p|}$ as well as $\DIST_p$.

The $\DIST_\edit$ problem is well known to be \NP-hard~\citep{grohe2018}, and is closely related to the \emph{maximum common edge subgraph problem}. In fact, the assignment $\pi$ minimising $\mu_\edit(G^\pi,H)$ minimises the volume of $G^\pi-H$, and maximises the number of neutral edges. The problems $\DIST_{|p|}$ and $\DIST_p$ have also been proven to be \NP-hard \citep{grohe_similarity}. As shown in the result below, the values of $\mu_p$ and $\mu_{|p|}$ are closely related to the \emph{maximum mismatch count} of assignments. In the language of the present paper, the maximum mismatch count of an assignment $\pi\in\Inj(G,H)$, denoted by $\MMC(\pi)$, is the maximum degree of the vertices in $G^\pi- H$.  

\begin{lem}[Lemmata~9 and 12,  \citep{grohe_similarity}]\label{lem:mmc}
Fix an assignment $\pi\in\Inj(G,H)$, and set $q=\MMC(\pi)$. For any $p\geq 1$ and $\Diamond\in \left\{p,|p|\right\}$, it holds that
\begin{equation}\label{eq:bound_mut}
\max\left\{q^{1/p},q^{1-1/p}\right\}\leq \mu_{\Diamond}(G^\pi,H)\leq q.
\end{equation}
\end{lem}
 
We remark that Lemma~\ref{lem:mmc} generalises some   well-known results from linear algebra and spectral graph theory. For instance, when $p=2$, the inequality of $\mu_{|2|}\left(G^\pi,H\right)\leq \MMC(\pi)$ is the restatement of the well-known fact that  the maximum eigenvalue of the adjacency matrix of a graph $G$ is at most the maximum degree of  $G$. On the other side, when $p=1$, Lemma~\ref{lem:mmc} implies that $\mu_{1}\left(G^\pi,H\right)=\mu_{|1|}\left(G^\pi,H\right)=\MMC(\pi)$, which can be shown from the well-known identity \begin{displaymath}
|\!|A|\!|_1=\max_{v\in V}\sum_{w\in V}\left| A_{wv}\right|.
\end{displaymath}

%Note that for $p=2$ the fact that $\mu_{|2|}\left(G^\pi,H\right)\leq \MMC(\pi)$ coincides with the well known fact that the largest eigenvalue of the adjacency matrix of a graph is bounded from above by the maximum degree of the graph. While not evident in the above statement, for $p=1$ it holds that $\mu_{1}\left(G^\pi,H\right)=\mu_{|1|}\left(G^\pi,H\right)=\MMC(\pi)$. This follows from the well known fact that for any $V\times V$ matrix $A$ $$ |\!|A|\!|_1=\max_{v\in V}\sum_{w\in V}\left| A_{wv}\right|. $$

\section{Computational Complexity of   the $\DIST_\edit$ Problem}\label{sec:volume}

This section studies the computational complexity of the $\DIST_\edit$ problem, and   is organised as follows. We construct the family of graphs used in our proof in Section~\ref{sec:GD_construction}, and employ these graphs to prove Theorem~\ref{thm:main_2} in Section~\ref{sec:diste_proof}.

\subsection{Construction of Input Instances \label{sec:GD_construction}}

The crux to understand mismatch norms for graphs is to find properties that have to be preserved by the alignments, which is usually   challenging for graphs with high symmetry or regularity. To overcome this,  we start with some graph $G$, and construct new graphs $G[q]$ and $D_{n,q}$ by attaching cliques of appropriate sizes to each vertex of $G$ and the $n$ vertex cycle respectively, so that the alignments must somewhat respect the cluster structure. For the remainder of this section, assume that $G$ is a $3$-regular graph.

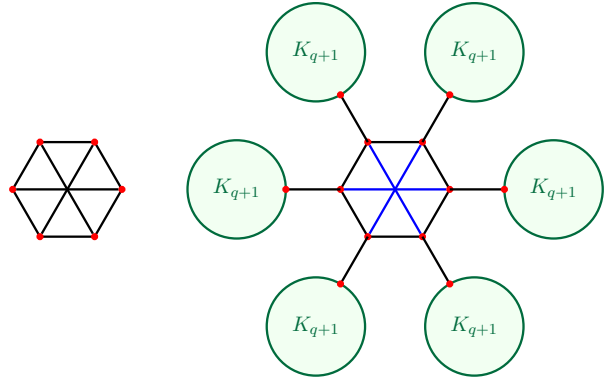
\begin{figure}[h]
    \centering
    \resizebox{8cm}{!}{  \begin{tikzpicture}[main/.style = {draw, fill, circle, inner sep=0pt, color=red, minimum size=2.5pt}][scale=2]
     
     \draw[very thick] (0.5,0.866) node[main] (1){} -- (1,0) node[main] (2){} -- (0.5,-0.866) node[main] (3){} -- (-0.5,-0.866) node[main] (4){} -- (-1,0) node[main] (5){} -- (-0.5,0.866) node[main] (6){} -- (0.5,0.866);

     \draw[very thick] (1)--(4);
     \draw[very thick] (2)--(5);
     \draw[very thick] (3)--(6);

     \draw[very thick] (6.5,0.866) node[main] (7){} -- (7,0) node[main] (8){} -- (6.5,-0.866) node[main] (9){} -- (5.5,-0.866) node[main] (10){} -- (5,0) node[main] (11){} -- (5.5,0.866) node[main] (12){} -- (6.5,0.866);

     \def\pointlist{(6.5,0.866), (7,0), (6.5,-0.866), (5.5,-0.866), (5,0), (5.5,0.866), (6.5,0.866)};

     \filldraw[color=newgreen, fill=green!5, very thick] (2.9*0.5+6,2.9*0.866) circle (0.9) node {$K_{q+1}$};
  \filldraw[color=newgreen, fill=green!5, very thick] (2.9+6,0) circle (0.9) node {$K_{q+1}$};
  \filldraw[color=newgreen, fill=green!5, very thick] (2.9*0.5+6,2.9*-0.866) circle (0.9) node {$K_{q+1}$};
  \filldraw[color=newgreen, fill=green!5, very thick] (2.9*-0.5+6,2.9*-0.866) circle (0.9) node {$K_{q+1}$};
  \filldraw[color=newgreen, fill=green!5, very thick] (2.9*-0.5+6,2.9*0.866) circle (0.9) node {$K_{q+1}$};
  \filldraw[color=newgreen, fill=green!5, very thick] (-2.9+6,0) circle (0.9) node {$K_{q+1}$};

  \draw[very thick] (0.5+6,0.866) -- (2*0.5+6,2*0.866) node  [main] (13){};
  \draw[very thick] (1+6,0) -- (2+6,0) node  [main] (14){};
  \draw[very thick] (0.5+6,-0.866) -- (2*0.5+6,-0.866*2) node  [main](15){};
  \draw[very thick] (-0.5+6,-0.866) -- (-1+6,-0.866*2) node [main](16){};
  \draw[very thick] (-1+6,0) -- (-2+6,0) node [main]  (17){};
  \draw[very thick] (-0.5+6,0.866) -- (-1+6,2*0.866) node [main]  (18){};

     \draw[color=blue, very thick] (7)--(10);
     \draw[color=blue, very thick] (8)--(11);
     \draw[color=blue, very thick] (9)--(12);
\end{tikzpicture}}
    \caption{The left is  a 3-regular graph $G$, and the right is our constructed $G\left[q\right]$. The green circles indicate the auxiliary $(q+1)$-cliques. }
    \label{fig:gq}
\end{figure}

\paragraph*{Construction of $G[q]$.} For any graph $G$ on $n$ vertices and a constant $q\in\mathbb{N}$,   we construct  the graph $G\left[q\right]$  as follows: 
\begin{itemize}
\item The vertex set of $G[q]$ is defined by 
\begin{displaymath}V\left(G\left[q\right]\right) := V(G)\cup \left\{ v^{\left[i\right]}\mid v\in V(G),\,1\leq i\leq q +1\right\};\end{displaymath}
\item The edge set of $G[q]$ is defined by 
\begin{displaymath}E\left(G\left[q \right]\right) := E(G)\cup \left\{\left(v^{\left[ 1\right]},v\right)\mid v\in V(G)\right\}\cup \left\{\left(v^{\left[i\right]}, v^{\left[j\right]}\right)\mid 1\leq i<j\leq q+1\right\}.
\end{displaymath}
\end{itemize}
That is, the vertices $v^{[i]}$ for $(1\leq i\leq q+1)$ form a $(q+1)$-clique, which is attached to the original graph $G$ via the edge $\left(v^{[1]},v \right)$; see Figure~\ref{fig:gq} for an illustration. In the following discussion, we refer to these cliques as \emph{auxiliary cliques}.

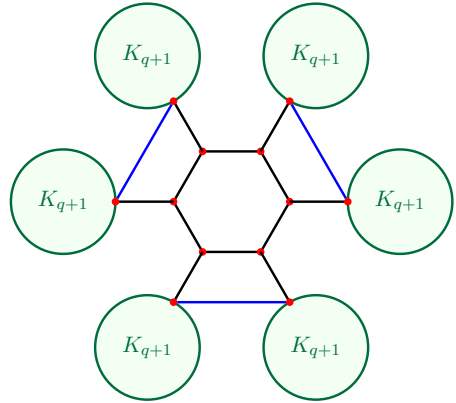
\begin{figure}[h!]
\centering
 \resizebox{6cm}{!}{
    \begin{tikzpicture}[main/.style = {draw, fill, circle, color=red, inner sep=0pt, minimum size=2.5pt}][scale=.2]
  \draw[very thick] (0.5,0.866) node[main] (1){} -- (1,0) node[main] (2){} -- (0.5,-0.866) node[main] (3){} -- (-0.5,-0.866) node[main] (4){} -- (-1,0) node[main] (5){} -- (-0.5,0.866) node[main] (6){} -- (0.5,0.866);

  \filldraw[color=newgreen, fill=green!5, very thick] (2.9*0.5,2.9*0.866) circle (0.9) node {$K_{q+1}$};
  \filldraw[color=newgreen, fill=green!5, very thick] (2.9,0) circle (0.9) node {$K_{q+1}$};
  \filldraw[color=newgreen, fill=green!5, very thick] (2.9*0.5,2.9*-0.866) circle (0.9) node {$K_{q+1}$};
  \filldraw[color=newgreen, fill=green!5, very thick] (2.9*-0.5,2.9*-0.866) circle (0.9) node {$K_{q+1}$};
  \filldraw[color=newgreen, fill=green!5, very thick] (2.9*-0.5,2.9*0.866) circle (0.9) node {$K_{q+1}$};
  \filldraw[color=newgreen, fill=green!5, very thick] (-2.9,0) circle (0.9) node {$K_{q+1}$};

  \draw[very thick] (0.5,0.866) -- (2*0.5,2*0.866) node  [main] (7){};
  \draw[very thick] (1,0) -- (2,0) node  [main] (8){};
  \draw[very thick] (0.5,-0.866) -- (2*0.5,-0.866*2) node  [main](9){};
  \draw[very thick] (-0.5,-0.866) -- (-1,-0.866*2) node [main](10){};
  \draw[very thick] (-1,0) -- (-2,0) node [main]  (11){};
  \draw[very thick] (-0.5,0.866) -- (-1,2*0.866) node [main]  (12){};

\draw[color=blue, very thick] (11)--(12);
\draw[color=blue, very thick] (9)--(10);
\draw[color=blue, very thick] (7)--(8);
\end{tikzpicture}}
\caption{The graph $D_{6,q}$ obtained by adding the blue edges to $\mathcal{C}_6\left[q\right]$.}
\label{fig:Dn}
\end{figure}

\paragraph*{Construction of $D_{n,q}$.} 
 For any even values of $n$ and $q\in\mathbb{N}$, we construct the graph $D_{n,q}$ as follows:
\begin{itemize}
\item The vertex set of $D_{n,q}$ is the same as that of $\C_n[q]$, where $\C_n$ is an $n$-cycle;
\item Let $M$ be a perfect matching of $\C_n$. The edge set of $D_{n,q}$ is defined as
\[
E(D_{n,q}) := E(\C_n[q]) \bigcup \left\{ \left(u^{[1]}, v^{[1]} \right) ~|~ (u,v)\in M \right\}.
\]
\end{itemize}
See Figure~\ref{fig:Dn} for an illustration.

%\red{To explain the intuition behind constructing $G[q]$ and $D_{n,q}$, one can view  $G$ and $\mathcal{C}_n$ inside $D_{n,q}$ as the   \emph{cores} of the two graphs, and the cliques connected to the cores will be used to significantly reduce the set of considered  alignments. On the other side, since these cliques help increase the size of input graphs, it indirectly help amplify the hardness of our studied problem. This will be clear when we analyse the reduction. }

It is important to notice that, since $\C_n$ has only two perfect matchings,   $D_{n,q}$ is uniquely defined up to the relabelling of the vertices. Moreover, by construction, graphs $G[q]$ and $D_{n,q}$ have the same number of edges.

\subsection{Proof of Theorem~\ref{thm:main_2}}\label{sec:diste_proof}

%Let $G$ be a $3$-regular graph on $n$ vertices and let $\C_n$ be the $n$-cycle. Recall the following `folklore' fact, from which the \NP-hardness of $\DIST_\edit$ immediately follows (the proof follows a reduction from the \NP-hard Hamiltonian cycle problem on $3$-regular graphs, analogously to Proposition \ref{prop:hardness_1}).

We first examine the core parts of  $G[q]$ and $D_{n,q}$~(i.e., $G$ and $\C_n$). Based on the \NP-hardness of the Hamiltonian cycle problem for 3-regular graphs \citep{akiyama}, it is easy to show that deciding whether $\delta_\edit(G,\C_n)\leq n/2$ is $\NP$-hard, thus implying the \NP-hardness of the  $\DIST_\edit$ problem.

\begin{fac}\label{fact:folklore}
It holds that $\delta_\edit(G,\C_n)\leq n/2$ if and only if  $G$ has a Hamiltonian cycle \footnote{This fact is usually deemed as `folklore'. A standard proof can be found in \citep{grohe2018}.}.
\end{fac}

The main part of the proof of Theorem \ref{thm:main_2} is to show that the alignments $\pi$ minimising $\mu_\edit\left(G\left[q\right]^\pi,D_{n,q}\right)$ must map the copy of the graph $G$ in $G\left[q\right]$ onto the copy of $\C_n$ in $D_{n,q}$. This allows us to consider only a restricted class of alignments in order to apply Fact \ref{fact:folklore}. Formally, we refer to these `restrictive alignments' as \emph{conservative alignments}.

\begin{defi}[Conservative alignment]
    Let $K$ and $L$ be subgraphs of $G$ and $H$, respectively. We say that $\pi\in\Inj(G,H)$ is $(K,L)$-\emph{conservative} if $\pi \mid _{V(K)}\in\Inj(K,L)$, where $\pi\mid_{V(K)}$ is the restriction of the alignment $\pi$ to the domain $V(K)\subseteq V(G)$.
\end{defi}

Put otherwise, for any graphs $G, H$  and their respective subgraphs $K, L$, we say that $\pi\in\Inj(G,H)$ is $(K,L)$-\emph{conservative} if the vertices of $K$ map to the vertices of $L$ under $\pi$. The following lemma presents a key property of conservative alignments: given a pair of $G[q], D_{n,q}$ constructed as above and some $\pi\in\Inj(G[q],D_{n,q})$, it shows that if $q$ is large enough, then the mismatch norm \begin{math}\mu_\edit\left(G\left[q\right]^\pi,D_{n,q}\right)\end{math} is minimised by taking $\pi$ to be a $\left(G,\C_n\right)$-conservative alignment of a particular form.

\begin{lem}
\label{lem:conservmapping}
Fix some graph $G$ on $n$ vertices, and let $G[q]$ and $D_{n,q}$ be graphs defined as above, where $q=3n+4$. Then \begin{displaymath}\mu_\edit\left(G\left[q\right]^\pi,D_{n,q}\right)=\delta_\edit\left(G\left[q\right],\C_n\right)\end{displaymath} 
holds only if $\pi\in\Inj\left(G\left[q\right],D_{n,q}\right)$ is $\left(G,\C_n\right)$-conservative. 
\end{lem}

\begin{proof} 
 We prove this by showing that if  $\pi$ is not $(G,\C_n)$-conservative,  then  $\mu_\edit(G[q]^\pi,D_{n,q})> 3n$; furthermore, there exists some $(G,\C_n)$-conservative mapping $\pi'$ such that $\mu_\edit(G[q]^{\pi'},D_{n,q})\leq 3n$, thus implying the required result.
 
 First notice that  any vertex of the form $v^{\left[i\right]}$ in $G\left[q\right]$ has degree at least $q=3n+4$, and every vertex of the copy of $\mathcal{C}_n$ in $D_{n,q}$ has degree $3$. If $\pi$ is not $\left(G,\mathcal{C}_n\right)$-conservative, we have
\begin{displaymath}\mu_\edit\left(G\left[q\right]^\pi,D_{n,q}\right)\geq\MMC(\pi)\geq 3n+4-3>3n\end{displaymath} 

as required. We next construct a $\left(G,\mathcal{C}_n\right)$-conservative alignment $\pi'$ so that
\begin{enumerate}
    \item the corresponding restriction $\pi'\mid_{V(G)}=\sigma\in\Inj(G,\C_n)$ minimises $\mu_\edit(G^\sigma,\C_n)$, and 
    \item the auxiliary cliques in $G\left[q\right]$ match with those of $D_{n,q}$; that is, $\pi'\left(v^{\left[ i\right]}\right)=\sigma(v)^{\left[i\right]}$ holds for all $v\in V(G)$ and $1\leq i\leq q+1$.
\end{enumerate}

As such, we have that \begin{math}\mu_\edit\left(G^\sigma,\C_n\right)=\delta_\edit\left(G,\C_n\right)\end{math}. Since $G$ is $3$-regular and $\C_n$ is $2$-regular, it follows that $G^\sigma-\C_n$ has at most $5$ incident edges per vertex, and hence \begin{displaymath}\mu_\edit\left(G\left[q\right]^{\pi'},D_{n,q}\right)=\delta_\edit(G,\C_n)+\frac{n}{2}\leq \frac{5n}{2}+\frac{n}{2}=3n.\end{displaymath}
This is because $\pi'$ as defined aligns all the auxiliary cliques, so the only edges not aligned are the ones in the mismatch graph $G^\sigma-\C_n$, whose number is at most $5n/2$, as well as $n/2$ edges corresponding to the perfect matching $M$ (the blue edges in Figure~\ref{fig:Dn}). 
\end{proof}

Combining Lemma~\ref{lem:conservmapping} with Fact~\ref{fact:folklore} gives us the following result.

\begin{lem}
\label{lem:amplificationone}
    Let $G$ be a 3-regular graph on $n$ vertices, and  $q=3n+4$. Then,  $\delta_\edit(G\left[q\right],D_{n,q})\leq n$ holds  if and only if $G$ has a Hamiltonian cycle.
\end{lem}

\begin{proof} 
 By Lemma~\ref{lem:conservmapping} and its proof, we know that \begin{math}\mu_\edit\left(G\left[q\right]^\pi,D_{n,q}\right)\end{math} achieves the minimum if  $\pi$ is $\left(G,\C_n\right)$-conservative and of the form
\begin{displaymath}
v\mapsto \sigma(v),
\end{displaymath}
\begin{displaymath}
v^{\left[i\right]}\mapsto \sigma(v)^{\left[i\right]},
\end{displaymath}
for all $v\in V(G)$ and $1\leq i\leq q+1$, where $\sigma\in \Inj\left(G,C_n\right)$ is the restriction of $\pi$ to $V(G)$. For such $\pi$, we choose $\sigma$ that minimises  $\mu_\edit\left(G^\sigma,\C_n\right)$, and have that 
\begin{displaymath}
\delta_\edit\left(G\left[q\right],D_{n,q}\right)=\mu_\edit\left(G\left[q\right]^\pi,D_{n,q}\right)=\mu_\edit\left(G^\sigma,\C_n\right)+\frac{n}{2}=\delta_\edit\left(G,\C_n\right)+\frac{n}{2}.
\end{displaymath}
By Fact~\ref{fact:folklore},  it holds that 
\begin{math}\delta_\edit\left(G,\C_n\right)\leq n/2\end{math} if and only if  $G$ has a Hamiltonian cycle. Therefore, \begin{math}\delta_\edit\left(G\left[q\right],D_{n,q}\right)\leq n\end{math} if and only if $G$ has a Hamiltonian cycle,
which proves the lemma.
\end{proof}

Notice that the size of input graphs in Lemma~\ref{lem:amplificationone} is a function of $n$ and $q$. By taking the orders of $n,q$ and $|V(G[q])|$ into account, we have the following hardness result, in which  we set $|V(G)|=|V(H)|=n$ for any pair of input graphs $G$ and $H$.

\begin{cor}
\label{cor:coramplificationone}
    There exist functions $t(n)=\Theta(n^{1/2})$ for which, given graphs $G$ and $H$ on $n$ vertices each and with the same number of edges, deciding whether $\delta_\edit(G,H)\leq t(n)$ is \NP-hard.
\end{cor}

\begin{proof} 
Let $G'$ be a 3-regular graph on $n'$ vertices,   $G:= G'[3n'+4]$, and $H:= D_{n',3n'+4}$.
     From Lemma \ref{lem:amplificationone}, $\delta_\edit(G,H)\leq n'$  holds if and only if $G'$ has a Hamiltonian cycle. Since $G$ and $H$ have the same number of edges and can be constructed from $G$ and $H$ in polynomial time respectively, the result follows by our choice of  $n'=\Theta(n^{1/2})$. 
\end{proof}

Finally, we sketch the proof of Theorem~\ref{thm:main_2}. Notice that the choice of   $q=3n+4$ in the analysis above was arbitrary and for the purpose of explaining a simplified version of the proof.    In fact, employing the same analysis  as for Lemma~\ref{lem:conservmapping},  it is not  difficult to  show that, for any $q\geq 3n+4$, the mismatch norm  $\mu_\edit\left(G\left[q\right]^\pi, D_{n,q}\right)$ achieves its minimum only if $\pi$ is $(G,\C_n)$-conservative. In particular,  one can choose $q=\lceil  n^r\rceil$ for any fixed $r>1$, and  any $\pi$ that is not $(G,\C_n)$-conservative would yield 
\begin{displaymath}
\mu_\edit\left(G\left[q\right]^\pi,D_{n,q}\right)\geq \MMC(\pi)\geq \left\lceil  n^{r}\right\rceil.
\end{displaymath}
On the other side, any   conservative $\pi$ with $\sigma=\pi\mid_{V(G)}$ and $\pi\left( v^{\left[i\right]}\right)=\sigma\left(v\right)^{\left[i\right]}$ gives us that
\begin{displaymath}
\mu_\edit\left(G\left[q\right]^\pi,D_{n,q}\right)= \mu_\edit\left(G^\sigma,C_n\right)+\frac{n}{2}.
\end{displaymath}
If $\sigma$ minimises \begin{math}\mu_\edit\left(G^\sigma,C_n\right)\end{math}, then \begin{math}\mu_\edit\left(G\left[q\right]^\pi,D_{n,q}\right)\leq 5n/2+n/2=3n\end{math} which is asymptotically less than $ n^{r}$. As such,  we have   the following stronger version   of Lemma~\ref{lem:amplificationone}.
\begin{lem}
 Let $t(n):= \Theta\left(n^r\right)$ with $r>1$, and $q:= \lceil t(n)\rceil$. Then, for any 3-regular graph $G$ on $n$ vertices, it holds that \begin{math}\delta_\edit\left(G\left[q\right],D_{n,q}\right)\leq n\end{math} if and only if $G$ has a Hamiltonian cycle.  
\end{lem}
Finally, since
 $G\left[q\right]$ has $n+n(q+1)=\Theta\left(n^{1+r}\right)$ vertices and both $G\left[q\right]$ and $D_{n,q}$ can be constructed from $G$ and $\C_n$ in polynomial time  respectively, Theorem \ref{thm:main_2} is directly implied by the following statement.
\begin{cor}\label{cor:cor_main_1}
For any function $t(n)=\Theta\left(n^{1/(1+r)}\right)$ with $r>1$, given any graphs $G$ and $H$ on $n$ vertices each and with the same number of edges, deciding whether $\delta_\edit(G,H)\leq t(n)$ is \NP-hard.  
\end{cor}

\section{Computational Complexity of   the $\DIST_p$ and $\DIST_{|p|}$ Problems}\label{sec:2ndmain}

This section studies  the computational complexity of the $\DIST_p$ and $\DIST_{|p|}$ problems, and   is organised as follows. We construct the family of graphs used in our proof in Section~\ref{sec:construction2}, and show some structural results thereof in Section \ref{sec:structural}. These are applied in the proof of Theorem~\ref{thm:main_1} which is split into the cases $p\neq2$ (Section \ref{sec:analysis2}) and $p=2$ (Section \ref{sec:analysis22}).

\subsection{Construction of Input Instances \label{sec:construction2}}

 Given some $k\in\mathbb{N}$, the construction starts with a 3-regular bipartite graph $G_k$ and a 3-regular non-bipartite graph $H_k$ with a Hamiltonian cycle. We further construct $\hat{G}_k$ as well as $\hat{H}_k$ based on $G_k$ and $H_k$.

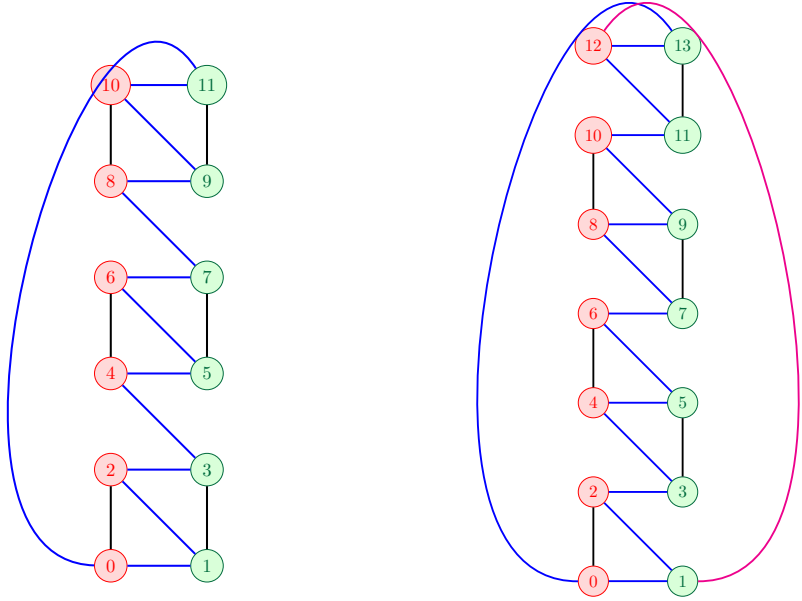
\begin{figure}[h!]
\centering
\hspace{-4cm}
     \parbox{1.2in}{ \resizebox{4.5cm}{!}{
    \begin{tikzpicture}[main/.style = {draw, circle, inner sep=0pt, minimum size=10pt}][scale=1.5]
    \foreach \i in {0,2,4,6,8,10}
    \node[draw,circle, color=red, fill=red!15] (\i) at (0,\i) {\i};

    \foreach \i in {1,3,5,7,9,11}
    \node[draw, circle, color=newgreen,fill=green!15] (\i) at (2,\i-1) {\i};

    \draw[color=blue, very thick] (0)--(1)--(2)--(3)--(4)--(5)--(6)--(7)--(8)--(9)--(10)--(11);
    \draw[color=blue, very thick] (0) to[bend left=100, in=45] (11);
    \draw[color=black, very thick] (0)--(2) (4)--(6) (8)--(10) (1)--(3) (5)--(7) (9)--(11);
    \end{tikzpicture}}

    \label{fig:2figsA}}%
  \hspace{3cm}
  \begin{minipage}{1.2in}%
   \resizebox{7.5cm}{!}{
\begin{tikzpicture}[main/.style = {draw, circle, inner sep=0pt, minimum size=10pt}][scale=2]
    \foreach \i in {0,2,4,6,8,10,12}
    \node[draw,circle, color=red, fill=red!15] (\i) at (0,\i) {\i};

    \foreach \i in {1,3,5,7,9,11,13}
    \node[draw, circle,   color=newgreen, fill=green!15] (\i) at (2,\i-1) {\i};

    \draw[color=blue, very thick] (0)--(1)--(2)--(3)--(4)--(5)--(6)--(7)--(8)--(9)--(10)--(11)--(12)--(13);
    \draw[color=blue, very thick] (0) to[bend left=100, in=45] (13);
    \draw[color=magenta, very thick] (1) to[bend right=100, in=315] (12);
    \draw[color=black, very thick] (0)--(2) (4)--(6) (8)--(10) (3)--(5) (7)--(9) (11)--(13);
    \end{tikzpicture}}
    \hspace{1cm}
    \label{fig:2figsB}%
  \end{minipage}%
  \caption{The left figure illustrates the construction of  $H_6$; notice that this is a planar graph.  The right figure illustrates the construction of $H_7$. Note that it is $1$-planar, for the edges $(0,13)$ and $(1,12)$ cross.\label{fig:hk}}
\end{figure}

\paragraph*{Construction of $G_k$ and $H_k$.}  For any $k\in\mathbb{N}$, let $G_k$ be a 3-regular bipartite graph on $n=2k$ vertices, and construct $H_k$ as follows:
\begin{itemize}
    \item The vertex set of $H_k$ is defined by
    $
    V(H_k) := \{0,1,\ldots, 2k-1\}$;
    \item The edge set of $H_k$ is defined by
    \[
    E(H_k) := E[\C_{2k}] \cup \left\{(i,j)\mid i\equiv0\,\mathrm{mod}\,4,\,j=i+2\right\}\cup\left\{(i,j)\mid i\equiv 1\,\mathrm{mod}\,4,\,j=i+2\right\}
    \]
    if $k$ is even, and
    \[
    E(H_k):= E[\C_{2k}]\cup\left\{(i,j)\mid i\equiv2\,\mathrm{mod}\,4,\,j=i-2\right\}\cup\left\{(i,j)\mid i\equiv 3\,\mathrm{mod}\,4,\,j=i+2\right\}\cup \left\{(2k-2, 1)\right\}
    \]
    if $k$ is odd. Here $\C_{2k}$ is the cycle graph of $2k$ vertices defined on $\{0,\ldots, 2k-1\}$.
\end{itemize}
See Figure~\ref{fig:hk} for an illustration of the construction of $H_k$.
 It's easy to see that $H_k$ is 3-regular, and the edges from $E[\C_{2k}]$ form a Hamiltonian cycle of $H_k$. Moreover, every vertex is part of a 3-cycle  if $k$ is even, and  every vertex except for $1$ and $2k-2$ is part of a 3-cycle if $k$ is odd. Thus, $H_k$ is not bipartite for any $k\in\mathbb{N}$.

\paragraph*{Construction of $\hat{G}_k$ and $\hat{H}_k$ from $G_k$ and $H_k$.}  For any given $G_k$, let $A,B\subset V(G_k)$ be equal-sized disjoint maximal independent sets of $V(G_k)$; notice that such $A,B$ always exist since $G_k$ is a regular bipartite graph. We construct $\hat{G}_k$ as follows:
\begin{itemize}
    \item Set $Q_A=\left\{ v^{\left[i\right]}\mid v\in A,\;1\leq i\leq 5\right\}$,  $Q_B=\left\{v^{\left[i\right]}\mid v\in B,\;1\leq i\leq 12\right\}$,
    and define \begin{displaymath}
V\left(\hat{G}_k\right):= V(G)\cup Q_A \cup Q_B;
\end{displaymath}
that is, we introduce 5 new neighbours $v^{\left[1\right]},\ldots v^{\left[5\right]}$ for every $v\in A$, and 12 new neighbours $v^{\left[1\right]},\ldots v^{\left[12\right]}$ for every $v\in B$.
\item The edge set of $\hat{G}_k$ is defined by
\[
E\left(\hat{G}_k\right):=E(G_k)\cup\left\{\left(v,v^{\left[i\right]}\right)\mid v\in A,\; 1\leq i\leq 5\right\}\cup \left\{\left(v,v^{\left[i\right]}\right)\mid v\in B,\; 1\leq i\leq 12\right\}; 
\]
that is, in addition to the existing edges from $G_k$, we add new edges to connect the vertices from $V(G_k)$ to their corresponding new neighbours. 
\end{itemize}

Our construction of $\hat{H}_k$ is similar to the one for $\hat{G}_k$: we partition $V\left(H_k\right)$ into sets of the same size defined by $\even\left(2k\right) := \left\{0,2,\hdots,2k-2\right\}$ and $\odd\left(2k\right) := \left\{1,3,\hdots,2k-1\right\}$. By setting \[
Q_{\even} := \left\{ v^{\left[i\right]}\mid v\in \even\left(2k\right),\;1\leq i\leq 5\right\}, \qquad Q_{\odd} := \left\{v^{\left[i\right]}\mid v\in \odd\left(2k\right),\;1\leq i\leq 12\right\},
\]
we define $V\left(\hat{H}_k\right)$ and $E\left(\hat{H}_k\right)$ in the same way as   $V\left(\hat{G}_k\right)$ and $E\left(\hat{G}_k\right)$, by replacing the sets $A$ and $B$ by $\even\left(2k\right)$ and $\odd\left(2k\right)$ respectively. It is important to note that whilst the construction of $\hat{G}_k$ depends on the choice of $A$ and $B$, one can \emph{uniquely} recover $G_k$ and $H_k$ from  $\hat{G}_k$ and $\hat{H}_k$ respectively in polynomial time.

At an intuitive level, these graphs   are  similar in spirit to the ones  from Section~\ref{sec:volume}. However, we attach `auxiliary leaves' (as opposed to cliques) to the vertices of the `old' graphs, so that the optimal conservative alignments are required to respect some partition of the vertices. In particular, we are able to guarantee that one only needs to consider $(G_k,H_k)$-conservative alignments $\pi$ for which $\pi(A)=\even\left(2k\right)$ and $\pi(B)=\odd\left(2k\right)$, which will be shown in the next section. This reduces the problem of minimising $\mu_p\left(G_k^\pi,H_k\right)$ over all $\pi$ to minimising $\mu_p\left(G_k^\sigma,H_k\right)$ over the alignments $\sigma$ respecting the partitions assigned to $G_k$ and $H_k$.

\subsection{Structural Results}\label{sec:structural}
In this section we list and prove some structural properties of the mismatch graph $\hat{G}_k^\pi-\hat{H}_k$. First, we show that, in order to minimise $\MMC\left(\pi\right)$, one only needs to consider alignments $\pi$ for which the edges incident to the `auxiliary leaves' disappear in the mismatch graph. Formally, any alignment $\pi\in \Inj\left(\hat{G}_k,\hat{H}_k\right)$ minimising $\MMC(\pi)$ is $\left(G_k,H_k\right)$-conservative, and  maps $A$ to $\even\left(2k\right)$ and $B$ to $\odd\left(2k\right)$, as shown in the following statement.

\begin{lem}\label{lem:mmc_conservative}
Let $\pi\in \Inj\left(\hat{G}_k,\hat{H}_k\right)$. The following hold:
\begin{enumerate}
    \item if $x\in Q_A\cup Q_B$ and $\pi(x)\in V\left(H_k\right)$, then $\MMC(\pi)>6$;
    \item if $x\in A$ and $\pi(x)\in \odd\left(2k\right)$, then $\MMC(\pi)>6$;
    \item assume $\pi$ is $\left(G_k,H_k\right)$-conservative, and that $\pi\left(A\right)=\even\left(2k\right)$ and $\pi\left(B\right)=\odd\left(2k\right)$. If $\sigma$ is the restriction of $\pi$ to $V(G_k)$ and $\pi$ satisfies $\pi\left(v^{\left[i\right]}\right)=\sigma(v)^{\left[i\right]}$ for all $v\in V(G_k)$ and suitable values of $i$, then $\MMC(\pi)=\MMC(\sigma)\leq 6$.
\end{enumerate}
\end{lem}

\begin{proof}
 To prove the first statement, let $x\in Q_A\cup Q_B$. Since $\pi(x)\in V\left(H_k\right)$, the degree of $\pi(x)$ in $\hat{H}_k$ is either 8 or 15. Hence, $\pi(x)$ has at least 7 more neighbours than $x$ if $\pi(x)\in \even\left(2k\right)$, and at least 14 more  neighbours if $\pi(x)\in \odd\left(2k\right)$.  This implies that  $\MMC(\pi)\geq 7>6$.

 For the second statement, since $x\in A$ and $\pi(x)\in \odd\left(2k\right)$, vertex $x$ has degree 8 and $\pi(x)$ in $\hat{H}_k$ has degree 15. This implies that $\MMC(\pi)\geq 7$.

 Finally, assume that $\pi$ satisfies the condition from the third statement. Then, the only edges appearing in $\hat{G}_k^\pi-\hat{H}_k$ are those of the mismatch graph $G_k^\sigma-H_k$. Since both of  $G_k$ and $H_k$ are $3$-regular,  we have that $\MMC(\pi)=\MMC(\sigma)\leq 6$.
\end{proof}

Lemma~\ref{lem:mmc_conservative} implies that, in order to minimise the maximum mismatch count, we only need to look at  the $\left(G_k,H_k\right)$-conservative alignments $\pi$ satisfying the conditions from the   third statement of the lemma.
 Indeed, suppose  \begin{math}\pi'\in \Inj\left(\hat{G}_k,\hat{H}_k\right)\end{math} is $\left(G_k,H_k\right)$-conservative, but \begin{math}\pi'\left(v^{\left[i\right]}\right)=w^{\left[j\right]}\end{math} for some $w\neq \pi'(v)$. We  consider the alignment $\pi$ defined as
\begin{displaymath}
v\mapsto \sigma(v)
\end{displaymath}
\begin{displaymath}
v^{\left[i\right]}\mapsto \sigma(v)^{\left[i\right]} 
\end{displaymath}
for all $v\in V(G)$ with $\sigma$ being the restriction of $\pi'$ to $V(G)$. Then, $\pi$ satisfies the conditions from the third statement of Lemma~\ref{lem:mmc_conservative}, and \begin{math}\MMC(\sigma)=\MMC(\pi)\leq \MMC\left(\pi'\right)\end{math}.

Taking this into account, let  \begin{math}\Res\left(G_k,H_k\right)\subset \Inj\left(G_k, H_k\right)\end{math}  be the set of alignments $\sigma$ with  $\sigma(A)=\even\left(2k\right)$ and $\sigma(B)=\odd\left(2k\right)$; that is,  every $\sigma\in \Res\left(G_k,H_k\right)$ is the restriction of some alignment $\pi$ satisfying the conditions from the third  statement of Lemma~\ref{lem:mmc_conservative}.

We now prove useful properties of the mismatch graph $G_k^\sigma-H_k$ for $\sigma\in\Res\left(G_k,H_k\right)$ \footnote{We note that the set of alignments $\Res\left(G_k,H_k\right)$ is defined only up to a choice of maximal independent sets $A$ and $B$. However, given a graph of the form $\hat{G}_k$ as above, the graph $G_k$ as well as a canonical choice of $A$ and $B$ are well defined by construction.}. The next three Lemmata effectively imply that computing $\MMC\left(\pi\right)$ is sufficient to decide whether or not $G_k$ has a Hamiltonian cycle.

\begin{lem}
\label{lem:hamiltoniandifference}
If $G_k$ has a Hamiltonian cycle, then there is some $\sigma\in \Res\left(G_k,H_k\right)$ for which $\MMC(\sigma)\leq 2$.
\end{lem}

\begin{proof} 
Notice that the vertices
$0, 1, 2,\ldots, 2k-1, 0$ form a Hamiltonian cycle of $H_k$. Taken in this order, this cycle alternates between the vertices from $\even\left(2k\right)$ and the ones from $\odd\left(2k\right)$. Hence, there exists some $\sigma\in\Res\left(G_k,H_k\right)$ for which $\sigma^{-1}(0), \sigma^{-1}(1),\ldots, \sigma^{-1}(2k-1), \sigma^{-1}(0)$ form a Hamiltonian cycle in $G_k$, implying that every  edge in the copy of $\C_{2k}$ in $H_k$ is neutral. Thus,  in the mismatch graph $G_k^\sigma-H_k$, each vertex has no neighbours, or connects to   exactly a positive incident edge and a negative one. This proves that  $\MMC(\sigma)\leq 2$.
\end{proof}

 \begin{lem}\label{lem:no_hamiltonian_difference_even}
Let $k$ be even. If $G_k$ has no Hamiltonian cycle, then it holds for any  $\sigma\in \Res(G_k,H_k)$ that $\MMC(\sigma)\geq 4$.
\end{lem}

\begin{proof}
 Assume ad absurdum that for some $\sigma$, all vertices in $G_k^\sigma-H_k$ have degree at most $2$. Since every vertex in $\even\left(2k\right)\subseteq V\left(H_k\right)$ is adjacent to another vertex of $\even\left(2k\right)$ and the vertices of $A\subseteq V(G)$ form an independent set, it follows that  every vertex of $\even\left(2k\right)$ in $G_k^\sigma-H_k$ connects to another  vertex of $\even\left(2k\right)$ via   a negative edge; a similar argument holds for $B$ and $\odd\left(2k\right)$.  Hence, by the assumption on $\sigma$ and  Fact \ref{fact:pos_neg}, each vertex of $G^\sigma_k-H_k$ has exactly one positive and one negative edge.

 Next, let
 \begin{displaymath}M:= \left\{(i,j)\mid i\equiv0\,\mathrm{mod}\,4,\,j=i+2\right\}\cup\left\{(i,j)\mid i\equiv 1\,\mathrm{mod}\,4,\,j=i+2\right\}\end{displaymath} 
 be the set of negative edges, which are coloured in black in  the left figure of Figure~\ref{fig:hk}. Hence, $E\left(H_k\right)\setminus M$ is the set of neutral edges under $\sigma$, and these edges exist both in $G_k^{\sigma}$ and $H_k$. Since $k$ is an even number, we have $E\left(H_k\right)\setminus M=\C_{2k}$, contradicting to the fact that $G_k$ doesn't have a Hamiltonian cycle. 
  Hence, some vertex  in $G_k^\sigma-H_k$ must have at least 2 incident negative edges and, by Fact \ref{fact:pos_neg}, the same vertex  must have at least 2 incident positive edges. This proves  that $\MMC(\sigma)\geq 4$.
\end{proof}

 For odd values of $k$, the analysis is similar though more involved, since there are two vertices of $H_k$ that are not part of any 3-cycle.

\begin{lem} 
\label{lem:nohamiltoniandifferenceodd}
Let $k$ be odd. If $G_k$ has no Hamiltonian cycle, then it holds for all $\sigma\in \Res\left(G_k, H_k\right)$ that  $\MMC(\sigma)\geq 4$.
\end{lem}

\begin{proof}
Our proof is similar to the one of   Lemma~\ref{lem:no_hamiltonian_difference_even}. We assume ad absurdum that there is some $\sigma\in\Res(G_k, H_k)$ for which all vertices in $G_k^\sigma-H_k$ have degree at most $2$. Applying the same argument as   in Lemma~\ref{lem:no_hamiltonian_difference_even}, we know that  every vertex from $A':= \even\left(2k\right)\setminus \left\{2k-2\right\}$ and $B':= \odd\left(2k\right)\setminus \left\{1\right\}$ connects  with another one of $A'$ and $B'$ through a negative edge, respectively. The edges in the set 
\[
M:=\left\{(i,j)\mid i\equiv2\,\mathrm{mod}\,4,\,j=i-2\right\}\cup\left\{(i,j)\mid i\equiv 3\,\mathrm{mod}\,4,\,j=i+2\right\},
\]
are therefore all negative. We continue the proof with the following case distinction: 
\begin{itemize}
    \item If the edge $(2k-2,1)$ is also negative, then every vertex in $G_k^\sigma-H_k$ has at least one negative edge, so $M\cup \left\{\left(2k-2,1\right)\right\}$ includes all edges that are negative for $\sigma$. This means that $E\left(H_k\right)\backslash \left(M\cup\left\{\left(2k-2,1\right)\right\}\right)=E\left(\C_{2k}\right)$, i.e., the neutral edges form a Hamiltonian cycle, which leads to a contradiction.
    \item Assume instead that edge $(2k-2,1)$ is neutral. Since $E\left(H_k\right)\setminus M=E\left(\C_{2k}\right)\cup \left\{\left(2k-2,1\right)\right\}$ and the edges of $\C_{2k}$ form a Hamiltonian cycle, it must be the case that $\C_{2k}$ includes some negative edge; we call such edge $(u,v)$.
    \begin{itemize}
        \item If $v\in A'\cup B'$, then $v$ has at least two incident negative edges: one is  from $M$, and the other is  $(u,v)\in E\left(\C_{2k}\right)$.
        \item Otherwise, we have $v\in\left\{1,2k-2\right\}$. Then, $u\notin \left\{1,2k-2\right\}$, since edge $(2k-2,1)$ is neutral by assumption. Hence, $u\in A'\cup B'$,  and $u$ has  at least two incident negative edges: one belongs to the set $M$,  and the other one is $(u,v)\in \C_{2k}$.
    \end{itemize}
    Combining the two cases with Fact~\ref{fact:pos_neg}, at least one vertex between $u$ and $v$ must   have at least two  incident positive edges as well, thus implying that $\MMC(\sigma)\geq 4$.
\end{itemize}
Combining the two cases together proves the lemma.
\end{proof}

Finally, we prove a technical result which is at the core of the proof of the case $p=2$ for Theorem \ref{thm:main_1}. While we formulate the Lemma in linear algebraic terms in order to apply it to Theorem \ref{thm:main_1}, it is effectively a statement on the combinatorial properties (indeed, the cycle structure) of $G_k^\sigma-H_k$.

\begin{lem}\label{lem:p=2}
    Let $A\left(\sigma\right)$ be the adjacency matrix of the mismatch graph $G_k^\sigma-H_k$ for some $\sigma\in\Res\left(G_k,H_k\right)$, and define $P\left(\sigma\right):= A\left(\sigma\right)^2$. Then for every vertex $x$ of degree $4$ in $G_k^\sigma-H_k$, there is a distinct vertex $v$ for which $P\left(\sigma\right)_{xv}\neq 0$.
\end{lem}

Let us discuss the above statement. Fix some $\sigma\in\Res\left(G_k,H_k\right)$, and define $P:= P\left(\sigma\right)$ and $A:= A\left(\sigma\right)$ as a shorthand. The reason for using $P$ rather than $A$ is the fact that the entries of $P$ have a neat combinatorial interpretation. Indeed, let $V$ be the vertex set of $G^\sigma_k-H_k$. Then for each vertex $x\in V$, the entry $P_{xx}$ is equal to the degree of $x$ in $G_k^\sigma-H_k$. More generally, for any $x,v \in V$, define $\mathcal{W}(x,v)$ to be the set of walks of length $2$ from $x$ to $v$ in $G_k^\sigma-H_k$. We define the \emph{sign} of a walk $W$, denoted $\sgn(W)$, to be the product of all the weights of the individual edges in the walk counting repetitions; so for example, the sign of the walk $W=w_1w_2w_3w_4$ in $G^\sigma_k-H$ is $A_{w_1w_2}A_{w_2w_3}A_{w_3w_4}$. In particular, the sign always takes value $\pm1$, since in our case the weights of the edges are always $\pm1$. We may now write $P_{xv}$ as follows:
\begin{displaymath}
P_{xv}=\sum_{W\in \mathcal{W}(x,v)} \sgn(W).
\end{displaymath}
This can be seen as a generalisation of the fact that the $(x,v)$ entry of the $k^{th}$ power of an adjacency matrix of an unweighted graph is given by the number of walks of length $k$ from $x$ to $v$ in said graph.

For the proof of Lemma~\ref{lem:p=2}, the reader may wish to observe the following properties about the mismatch graph $G^\sigma_k-H_k$, which can be easily deduced from the proofs of Lemmata \ref{lem:no_hamiltonian_difference_even} and \ref{lem:nohamiltoniandifferenceodd}. Recall that, by the definition of $G_k^\sigma-H_k$, it holds that $V=V\left(H_k\right)=\left\{0,1,\hdots,2k-1\right\}$, and that we have partitioned $V$ into the subsets $\even\left(2k\right)$ and $\odd\left(2k\right)$. Thus, we may assign a \emph{parity} to each vertex in the natural way.

\begin{pro}\label{pro:aux}
     For any even number $k$ and any $\sigma\in\Res\left(G_k,H_k\right)$, the following holds for $G_k^\sigma-H_k$: each vertex is adjacent to exactly one other vertex of the same parity via a negative edge, and has degree equal to $2$, $4$ or $6$.
\end{pro}
\begin{proof}
    By the definition of $\Res\left(G_k,H_k\right)$, the bijection $\sigma$ maps independent sets $A$, $B$ of size $k$ to $\even\left(2k\right)$ and $\odd\left(2k\right)$ respectively. The subgraphs induced by $\even\left(2k\right)$ and $\odd\left(2k\right)$ in $H_k$ are matchings. Since $A$ and $B$ are independent sets, the subgraphs induced by $\even\left(2k\right)$ and $\odd\left(2k\right)$ in $G_k^\sigma-H_k$ are also matchings, with each edge acquiring a negative sign. Thus, each vertex of $G_k^\sigma-H_k$ acquires exactly a neighbour of the same parity via a negative edge. By Fact \ref{fact:pos_neg}, each vertex must have even degree no larger than $6$. Since each vertex has at least one incident edge, there cannot be any vertices of degree $0$; hence, the remainder of the required statement follows.
\end{proof}

\begin{pro}\label{pro:aux_odd}
    For any odd number $k$ and any $\sigma\in\Res\left(G_k,H_k\right)$, the following hold for any vertex $x$ of $G^\sigma_k-H_k$:
    \begin{enumerate}
        \item If $x\notin\left\{1,2k-2\right\}$, then $x$ is adjacent to exactly one other vertex of the same parity via a negative edge, and has degree equal to $2$, $4$ or $6$..
        \item If $x\in \left\{1,2k-2\right\}$, then $x$ has degree $0,2,4$ or $6$ and all its neighbours have opposite parity.
    \end{enumerate}
\end{pro}
\begin{proof}
    By the definition of $\Res\left(G_k,H_k\right)$, the bijection $\sigma$ maps independent sets $A$, $B$ of size $k$ to $\even\left(2k\right)$ and $\odd\left(2k\right)$ respectively. In the graph $H_k$, all vertices but $1$ and $2k-2$ have exactly one neighbour of the same parity. Combining this with Fact \ref{fact:pos_neg} yields the first statement with a similar argument as in Proposition \ref{pro:aux}.

    For the second statement, note that the vertices $1$ and $2k-2$ have no neighbours of the same parity. So in the mismatch graph $G_k^\sigma-H_k$ either they are isolated or they have an even number of neighbours of the opposite parity by Fact \ref{fact:pos_neg}.
\end{proof}

\begin{figure}[h!]
\centering
\hspace{-1cm}
     \parbox{1.2in}{ \resizebox{3.5cm}{!}{
     \begin{tikzpicture}[main/.style = {draw, circle, inner sep=0pt, minimum size=10pt}][scale=1]
   
    \node[draw,circle, color=red, fill=red!15] (x) at (-2,0) {$x$};
    \node[draw,circle, color=red, fill=red!15] (z) at (-2,-6) {$z$};
    \node[draw,circle, color=green, fill=green!15] (w_1) at (1.5,0) {$w_1$};
    \node[draw,circle, color=green, fill=green!15] (v_1) at (1.5,-2) {$v_1$};
    \node[draw,circle, color=green, fill=green!15] (w_2) at (1.5,-4) {$w_2$};
    \node[draw,circle, color=green, fill=green!15] (v_2) at (1.5,-6) {$v_2$};
    \node[draw,circle, color=green, fill=green!15] (w_3) at (1.5,-8) {$w_3$};
    \node[draw,circle, color=green, fill=green!15] (v_3) at (1.5,-10) {$v_3$};
    \node[draw,circle, color=green, fill=green!15] (v_4) at (1.5,-12) {$v_4$};

    \draw[color=black, thick] (x)--(z) (x)--(w_1) (w_1)--(v_1) (w_2)--(v_2) (w_3)--(v_3) (z)--(v_2);
    \draw[color=blue, thick] (x)--(w_2) (x)--(w_3) (z)--(v_1) (z)--(v_3);
    \draw[color=blue, thin, dash pattern={on 7pt off 2pt on 1pt off 3pt}] (v_4)--(z);
    \draw[color=black, thin, dash pattern={on 7pt off 2pt on 1pt off 3pt}] (1.5,-11)--(z);
    
    \end{tikzpicture}
    }

    }%
  \hspace{5cm}
  \begin{minipage}{1.2in}%
   \resizebox{4cm}{!}{
   \begin{tikzpicture}[main/.style = {draw, circle, inner sep=0pt, minimum size=10pt}][scale=1]
   
    \node[draw,circle, color=red, fill=red!15] (x) at (-2,0) {$x$};
    \node[draw,circle, color=red, fill=red!15] (z) at (-2,-6) {$z$};
    \node[draw,circle, color=green, fill=green!15] (1) at (1.5,0) {$1$};
    
    \node[draw,circle, color=green, fill=green!15] (w_1) at (1.5,-2) {$w_1$};
    \node[draw,circle, color=green, fill=green!15] (v_1) at (1.5,-4) {$v_1$};
    \node[draw,circle, color=green, fill=green!15] (w_2) at (1.5,-6) {$w_2$};
    \node[draw,circle, color=green, fill=green!15] (v_2) at (1.5,-8) {$v_2$};
    \node[draw,circle, color=green, fill=green!15] (v_3) at (1.5,-10) {$v_3$};

    \draw[color=black, thick] (x)--(z) (x)--(1) (w_1)--(v_1) (w_2)--(v_2) (z)--(v_2);
    \draw[color=blue, thick] (x)--(w_1) (x)--(w_2) (z)--(v_1) (z)--(v_3);
    \draw[color=blue, thin, dash pattern={on 7pt off 2pt on 1pt off 3pt}] (1)--(z);
    \end{tikzpicture}
}
    \hspace{1cm}
    
  \end{minipage}%
  \caption{The left figure is used in the  proof of Lemma~\ref{lem:p=2} for even valued $k$, and  the dashed edges represent the case for which $x$ has degree $6$. The right figure is used in the  proof for odd valued $k$ for the case where $x\neq 2k-2$ is a neighbour of $1$ (the case $v_3=1$ is represented by the dashed \textcolor{blue}{blue} edge), and  negative edges are indicated in black and positive edges in \textcolor{blue}{blue}.}\label{fig:lemmap=2}
\end{figure}
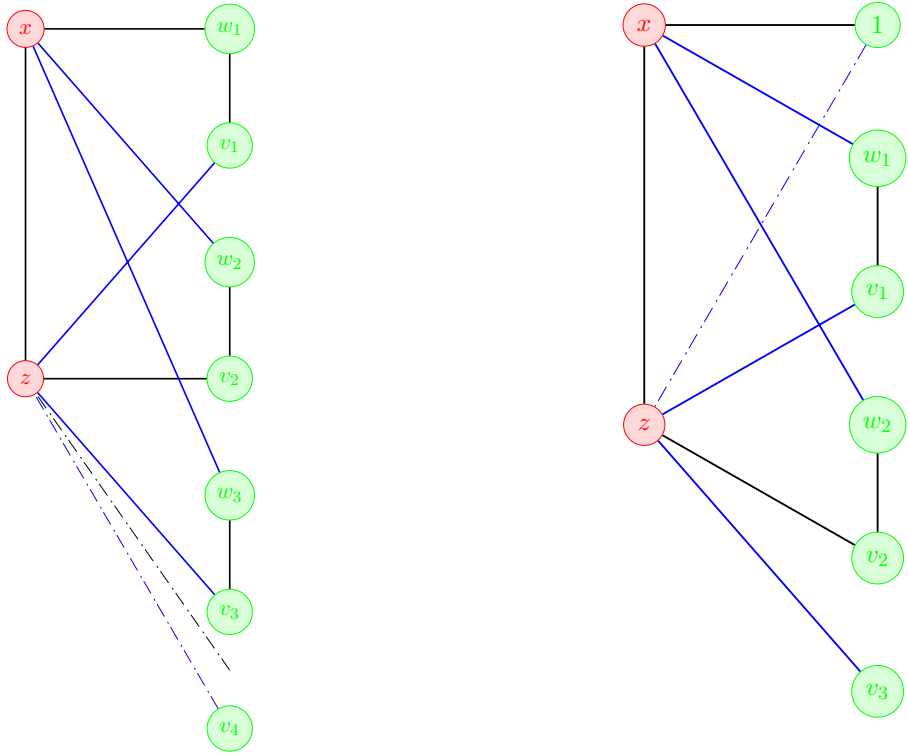

We may now prove Lemma \ref{lem:p=2}. Effectively, what we show is that for every vertex $x$ of degree $4$ there is a vertex $y$ of the opposite parity for which $\mathcal{W}\left(x,y\right)$ is non-empty,  and either $\left|\mathcal{W}\left(x,y\right)\right|=1$, in which case $P_{xy}=\pm1$, or $\left|\mathcal{W}\left(x,y\right)\right|=2$ and both walks have the same sign (equivalently, $x$ is part of a $4$-cycle with sign $+1$), thus implying that $P_{xy}=\pm2$. The reader may wish to consult Figure \ref{fig:lemmap=2} to aid the proof.

\begin{proof}[of Lemma \ref{lem:p=2}]

Fix some vertex $x$ of degree $4$, and assume without loss of generality that $x\in\even\left(2k\right)$.

     We first prove the statement for even valued $k$. By Proposition \ref{pro:aux}, $x$ has three odd neighbours, so there are exactly three walks of length $2$, say $K_1,K_2,K_3$, starting from $x$, going through an odd vertex and ending in some other odd vertex. We may write $K_1,K_2,K_3$ explicitly as follows:
        \begin{displaymath}
        K_1=xw_1v_1 
        \end{displaymath}
        \begin{displaymath}
        K_2=xw_2v_2
        \end{displaymath}
        \begin{displaymath}
        K_3=xw_3v_3
        \end{displaymath}
        where $w_i,v_i\in\odd\left(2k\right)$ for $i\leq 3$ and, by Proposition \ref{pro:aux}, the edges $\left(w_i,v_i\right)$ are negative since they join vertices of the same parity. If there is some $i\leq 3$ for which $K_i$ is the \emph{unique} element of $\mathcal{W}\left(x,v_i\right)$, then we are done, for this would imply that $P_{xv_i}=\pm1$. Otherwise, there is some $z\in\even\left(2k\right)$ such that each $v_i$ for $i\leq 3$ is a neighbour of $z$ and $\left(x,z\right)$ is a negative edge. In this case, we have that $\left|\mathcal{W}\left(x,v_i\right)\right|=2$ for each $i\leq 3$, for in addition to $K_i$, $v_i$ can be reached in two steps via the walk $L_i=xzv_i$.
        
        Since $z$ has at least $3$ distinct odd neighbours, namely $v_1,v_2,v_3$, then it must have degree $4$ or $6$ by Proposition \ref{pro:aux}. If $z$ has degree $6$, then it must have an odd neighbour $v_4\notin\left\{v_1,v_2,v_3\right\}$. Since $x$ has neighbourhood $\left\{z,w_1,w_2,w_3\right\}$, and $v_4$ is not a neighbour of $w_1,w_2$ or $w_3$, the only walk of length $2$ from $x$ to $v_4$ is $xzv_4$, thus implying $P_{xv_4}=\pm1$.
        
        If instead, $z$ has degree $4$, note that exactly one of the paths $K_1,K_2,K_3$ has positive sign. Similarly, exactly one of the paths $L_1=xzv_1, L_2=xzv_2,L_3=xzv_3$ has positive sign. Hence, there must be some $i\leq 3$ for which both $K_i$ and $L_i$ have the same sign. This gives at least two walks of length $2$ from $x$ to $v_i$, both with the same sign; whence, $P_{xv_i}=\pm2$ as required.

    We now prove the lemma for odd valued $k$ by distinguishing the following three cases:
    \begin{enumerate}
        \item \textbf{The vertex $x\neq 2k-2$ is not a neighbour of $1$}. In this case, we may argue similarly to the case where $k$ is even.

        \item \textbf{The vertex $x\neq 2k-2$ is a neighbour of $1$}. Since $1$ is a neighbour of $x$ and has no odd neighbours by Proposition \ref{pro:aux_odd}, there are exactly two distinct walks of length $2$ from $x$ to an odd vertex passing through an odd vertex. Write these walks as $xw_1v_1$ and $xw_2v_2$ where $w_i,v_i\in \odd\left(2k\right)$ for $i\leq 2$. If either of this is the unique element of $\mathcal{W}\left(x,v_i\right)$ for their respective value of $i$, then we are done, for it implies that $P_{xv_i}=\pm1$. Otherwise, there is some even neighbour of $x$, say $z$, such that $xzv_1$ and $xzv_2$ are distinct walks of length $2$ in $G_k^\sigma-H_k$. Applying Proposition \ref{pro:aux_odd}, we note that $z\neq 2k-2$, as it is a neighbour of the even vertex $x$. Since $z$ has two odd neighbours, it must have degree at least $4$ and thus a further odd neighbour $v_3\notin\left\{v_1,v_2\right\}$. We now distinguish between the cases where $v_3=1$ and $v_3\neq 1$:

        \begin{itemize}
            \item If $v_3=1$, then the walk $xzv_3$ is the unique walk of length $2$ from $x$ to $v_3$, thus $P_{xv_3}=\pm1$.
            \item If $v_3\neq 1$, then by construction $v_3$ cannot be reached by $x$ via a walk of length $2$ other than $xzv_3$. This is because the odd neighbours of $x$ are $1,w_1$ and $w_2$, and the only odd vertices adjacent to these are $v_1$ and $v_2$ (recall from Proposition \ref{pro:aux_odd} that vertices have at most one neighbour of the same parity). This means that $xw'v_3$ is the unique element of $\mathcal{W}\left(x,v_3\right)$ and therefore $P_{xv_3}=\pm1$ as required.
        \end{itemize}

        \item \textbf{Finally, $x=2k-2$}. In this case, Proposition \ref{pro:aux_odd} implies that since $x$ has degree $4$, it must have an odd neighbour $w\neq 1$. Since $w\neq 1$, it follows that $w$ has an odd neighbour $v$. Since $x$ has no even neighbours, we have that $xwv$ is the only walk of length $2$ from $x$ to $v$ so $P_{xv}=\pm 1$ as required.
    \end{enumerate}
\end{proof}

\subsection{Proof of Theorem~\ref{thm:main_1} for $p\neq 2$}\label{sec:analysis2}

As previously mentioned, computing the minimum value of $\MMC\left(\sigma\right)$ over all $\sigma\in\Res\left(G_k,H_k\right)$ suffices to determine whether $G_k$ has a Hamiltonian cycle. Formally, we have the following.

 \begin{pro}
     \label{prop:hardness_B}
     The following conditions are equivalent:
    \begin{enumerate}
        \item $G_k$ has a Hamiltonian cycle;
        \item $\min_{\sigma\in\Res\left(G_k,H_k\right)}\MMC(\sigma)\leq 2$;
        \item $\delta_1\left(\hat{G}_k,\hat{H}_k\right)=\delta_{|1|}\left(\hat{G}_k,\hat{H}_k\right)\leq 2$.
    \end{enumerate}
 \end{pro}

\begin{proof}
 The equivalence between (1) and (2) follows from Lemmata~\ref{lem:hamiltoniandifference}, \ref{lem:no_hamiltonian_difference_even} and \ref{lem:nohamiltoniandifferenceodd}.  The equivalence between (2) and (3) follows by the fact that we can take an optimal alignment $\pi\in\Inj\left(\hat{G}_k,\hat{H}_k\right)$ to satisfy the conditions from the third statement of Lemma~\ref{lem:mmc_conservative} and the fact that, for any graphs $G, H$ and alignment $\pi\in\Inj(G,H)$, it holds that $\mu_1(G^\pi,H)=\mu_{|1|}(G^\pi,H)=\MMC(\pi)$.
\end{proof}

 Now we are ready to prove Theorem~\ref{thm:main_1} for the case $p\neq 2$.
 
\begin{proof}[of Theorem~\ref{thm:main_1} for $p\neq 2$]
 Let $\mathcal{B}$ be the class of 1-planar graphs with degree at most 15. By construction, $\hat{G}_k$ and $\hat{H}_k$ have the same degree sequence. Let $G_k$ be planar, and observe from the drawings that $H_k$ is $1$-planar, so by construction $\hat{G}_k,\hat{H}_k\in\mathcal{B}$. Notice that $G_k$ is defined to be a 3-regular planar bipartite graph, so it is \NP-hard to decide whether  $G_k$ has a Hamiltonian cycle~\citep{akiyama}.
 
We first consider the case $p=1$. By Proposition~\ref{prop:hardness_B}, we know that deciding whether  $G_k$ has a Hamiltonian cycle is equivalent to checking whether there is some $\pi\in\Inj\left(\hat{G},\hat{H}_k\right)$ for which $\MMC(\pi)\leq 2$. Since $\hat{G}_k$ and $\hat{H}_k$ can be constructed in polynomial time from $G_k$ and $H_k$ respectively,  the problems $\DIST_1$ and $\DIST_{|1|}$ are \NP-hard for any  pair of graphs from $\mathcal{B}$ with the same degree sequence.

%Recall that it is \NP-hard to decide whether a $3$-regular planar bipartite graph has a Hamiltonian cycle~\he{add ref.}. Proposition \ref{prop:hardness_2} show that there is a polynomial-time reduction from this problem to the problem of deciding whether there is some $\pi\in\Inj\left(\hat{G},\hat{H}_k\right)$ for which $\MMC(\pi)\leq 2$. Thus $\DIST_1$ and $\DIST_{|1|}$ are \NP-hard for pairs of graphs from $\mathcal{B}$ with the same degree sequence.

%From Lemma \ref{lem:mmc} it follows that $\mu_2\left(\hat{G}^\pi,\hat{H}_k\right)\leq 2$ if $G$ has a Hamiltonian cycle. If not, then the bottom inequality and Proposition \ref{prop:hardness_2} imply that $\mu_2\left(\hat{G}^\pi,\hat{H}_k\right)>\sqrt{4}= 2$. So $\DIST_2$ is \NP-hard for pairs of graphs from $\mathcal{B}$ with the same degree sequence and a similar argument applies to $\DIST_{|2|}$.
Next, we consider the case $p>2$. By Lemma \ref{lem:mmc}, we have  $\mu_p\left(\hat{G}^\pi_k,\hat{H}_k\right)\leq 2$ if $G_k$ has a Hamiltonian cycle, and $\mu_p\left(\hat{G}^\pi_k,\hat{H}_k\right)\geq4^{1-1/p}>2$ otherwise. A similar conclusion to the above then follows for $\DIST_p$ and $\DIST_{|p|}$ for $p>2$.
\end{proof}

\subsection{Proof of Theorem \ref{thm:main_1} for $p=2$}\label{sec:analysis22}
The above argument for the reduction from the Hamiltonian cycle problem on $3$-regular graphs to $\DIST_p$ using Lemma \ref{lem:mmc} clearly fails if $p=2$. A priori, it does not exclude the existence of some $G_k$ and $\sigma\in\Res\left(G_k,H_k\right)$ for which the mismatch graph $G_k^\sigma-H_k$ has largest degree equal to $4$ and $\mu_2\left(G_k^\sigma,H_k\right)=\sqrt{4}=2$. If this were the case, then the metrics $\delta_2$ and $\delta_{|2|}$ would not be able to distinguish instances where $G_k$ has a Hamiltonian cycle or otherwise. However, it turns out that for any $\sigma\in\Res\left(G_k,H)k\right)$ for which $G_k^\sigma-H_k$ has a vertex of degree larger than $4$, the equality $\mu_2\left(G_k^\sigma,H_k\right)=\sqrt{4}=2$ can never occur, as shown in the following statement.

\begin{lem}\label{lem:matrix_P}
    Let $A\left(\sigma\right)$ be the adjacency matrix of the mismatch graph $G_k^\sigma-H_k$ for some $\sigma\in\Res\left(G_k,H_k\right)$, and define \begin{math}P\left(\sigma\right):= A\left(\sigma\right)^2\end{math}. If $G_k^\sigma-H_k$ has a vertex of degree at least $4$, then it holds that \begin{math}\left|\!\left|P\left(\sigma\right)\right|\!\right|_2=\left|\!\left|A\left(\sigma\right)\right|\!\right|_2^2>4\end{math}.
\end{lem}

\begin{proof} Set $P:= P\left(\sigma\right)$ for brevity. First of all, we assume that 
 $G_k^\sigma-H_k$ has maximum degree equal to $6$. Then, Lemma~\ref{lem:mmc} implies that $$\mu_{2}\left(G_k^\sigma,H_k\right)=\left|\!\left|A\left(\sigma\right)\right|\!\right|_2\geq\sqrt{6}>2$$ as required.
    
   Secondly, we assume that  $G_k^\sigma-H_k$ has maximum degree equal to $4$. Let $x$ be a vertex of degree $4$, and thus $P_{xx}=4$. By Lemma \ref{lem:p=2}, there must be some $v\neq x$ for which $P_{xv}\neq 0$. Consider the $2\times 2$ matrix $P'$ defined as
    \begin{displaymath}
    P':=\begin{pmatrix}
        P_{xx} & P_{xv}\\
        P_{vx} & P_{vv}
    \end{pmatrix}.
    \end{displaymath}
    Note that $P_{xv}\neq 0$, so it must be that $P_{vv}\geq 2$ since the degree of each vertex in $G^\sigma_k-H_k$ is even by Fact \ref{fact:pos_neg}. The eigenvalues of $P'$ are the roots of the quadratic polynomial
    \begin{displaymath}
    \mathrm{det}\left(tI-P'\right)=\left(t-P_{xx}\right)\left(t-P_{vv}\right)-P_{xv}^2,
    \end{displaymath}
    where we have used the fact that $P$ is symmetric and thus $P_{xv}=P_{vx}$. Using the fact that $P_{xx}=4$ and $P_{vv}\in\left\{2,4\right\}$, it is easy to check that the above quadratic equation always has a root $\lambda>4$. Let $z\in\mathbb{R}^2$ be a unit eigenvector of $P'$ associated with the eigenvalue $\lambda$. Define $\hat{z}\in\mathbb{R}^V$ to be the column vector for which $\hat{z}_x=z_1$, $\hat{z}_v=z_2$, with the remaining entries being set to $0$. We then have
    \begin{displaymath}
    \left|\!\left|P\right|\!\right|_2\geq\hat{z}^TP\hat{z}=z^TP'z=\lambda>4
    \end{displaymath}
    as required.
\end{proof}

The main consequence of the above is the following statement.

\begin{pro}
\label{pro:hardnessthree}
    The following are equivalent:
    \begin{enumerate}
        \item $G_k$ has a Hamiltonian cycle.
        \item \begin{math}\min_{\sigma\in\Res\left(G_k,H_k\right)}\mu_{|2|}\left(G_k^\sigma,H_k\right)\leq 2\end{math}.
        \item \begin{math}\min_{\sigma\in\Res\left(G_k,H_k\right)}\mu_{2}\left(G_k^\sigma,H_k\right)\leq 2\end{math}.
    \end{enumerate}
\end{pro}
\begin{proof}
    If $G_k$ has a Hamiltonian cycle, then by Lemma \ref{lem:hamiltoniandifference} there is some $\sigma\in\Res\left(G_k,H_k\right)$ for which all vertices in $G_k^\sigma-H_k$ have degree at most $2$. By Lemma \ref{lem:mmc}, it follows that \begin{math}\delta_2\left(G_k,H_k\right)\leq\mu_2\left(G_k^\sigma,H_k\right)\leq 2\end{math}, thus showing that (1)$\implies$(3). A similar argument shows that (1)$\implies$(2).

Next, we prove that   (2)$\implies$(1). We  assume that $G_k$ has no Hamiltonian cycle, fix some $\sigma\in\Res\left(G_k,H_k\right)$ and let $\lambda$ be the largest eigenvalue of the adjacency matrix of the \emph{unsigned} version of $G^\sigma_k-H_k$. By Lemmata~\ref{lem:no_hamiltonian_difference_even} and \ref{lem:nohamiltoniandifferenceodd}, some connected component of $G^\sigma_k-H_k$ has vertices with minimum degree at least $2$ and maximum degree at least $4$. Since $\lambda$ is at least as large as the average degree, it follows that $\mu_{|2|}\left(G_k^\sigma,H_k\right)=\lambda>2$ as required.  

    Finally, we prove (3)$\implies$(1). Assume that $G_k$ has no Hamiltonian cycle. From Lemmata \ref{lem:no_hamiltonian_difference_even} and \ref{lem:nohamiltoniandifferenceodd} it follows that for any $\sigma\in\Res\left(G_k,H_k\right)$, the mismatch graph $G_k^\sigma-H_k$ has a vertex of degree at least $4$. Applying Lemma \ref{lem:matrix_P}, we then get that for any $\sigma\in\Res\left(G_k,H_k\right)$ it holds that $\mu_2\left(G_k^\sigma,H_k\right)>2$ as desired.
\end{proof}

 We may now prove the case $p=2$ for Theorem \ref{thm:main_1}.

\begin{proof}[of Theorem~\ref{thm:main_1} for $p=2$]
    The statement follows using Proposition \ref{pro:hardnessthree} and an analogous argument to the proof of Theorem \ref{thm:main_1} for $p\neq 2$ as above.
\end{proof}

We finally remark that a priori, $\hat{G}_k$ and $\hat{H}_k$ as constructed do not imply a \NP-hardness result for the $\DIST_\edit$ problem, for it is unclear whether the required global properties of $G_k^{\sigma}-H_k$ can be deduced from Lemmata~\ref{lem:no_hamiltonian_difference_even} and \ref{lem:nohamiltoniandifferenceodd}. Indeed, we know that
\begin{itemize}
    \item if $G_k$ has a Hamiltonian cycle, then by  Lemma~\ref{lem:hamiltoniandifference} there is some $\sigma\in\Res\left(G_k,H_k\right)$ for which $G_k^\sigma-H_k$ is a disjoint union of cycles and has therefore no more than $n$ edges, which implies that $\delta_\edit\left(G_k,H_k\right)\leq n$;
    \item if $G_k$ has no Hamiltonian cycle and $k$ is even,  then Lemma~\ref{lem:no_hamiltonian_difference_even} ensures that $G_k^\sigma-H_k$ has at least two edges per vertex, and at least one vertex with 4 incident edges, giving a total of at least $\left(2(n-1)+4\right)/2=n+1$ edges;
    \item if $G_k$ has no Hamiltonian cycle and $k$ is odd, the best lower bound for $\mu_\edit\left(G_k^\sigma,H_k\right)$ is $n$, since Lemma~\ref{lem:nohamiltoniandifferenceodd} ensures the existence of only $n-2$ vertices of $G_k^\sigma-H_k$ with degree at least 2, and a vertex of degree 4, thus giving a lower bound of $\left(2(n-2)+4\right)/2=n$ edges. This is due to the fact that the vertices $1$ and $2k-2$ of $H_k$ are not part of any 3-cycles. 
\end{itemize}
The difference is meager, so it would not come as a surprise if a more careful analysis of $G_k^\sigma- H_k$ or a slightly different construction would yield a hardness result for $\DIST_\edit$ on pairs of graphs with same degree sequence. Nevertheless, the arguments and the constructions used to show Theorem~\ref{thm:main_1} seem to suggest some intrinsic differences between the nature and hardness of edit distance and the metrics associated with operator norms for different values of $p$. More precisely, the structural discrepancy between $\hat{G}_k$ and $\hat{H}_k$ for odd values of $k$ seems to be more easily detectable by local measures of similarity. For $\DIST_2$, this detectable structural discrepancy can be understood more concretely in terms of the cycle structure in the mismatch graph --- the main property exploited in the proof of Lemma \ref{lem:p=2}.

\section{Similarity Problems for Strongly Regular Graphs\label{sec:srg}}

This section studies the similarity problems for strongly regular graphs, and proves Theorem~\ref{thm:main_3}. The section is organised as follows: we  present the basics of strongly regular graphs in Section~\ref{sec:basics}, and the proof of Theorem~\ref{thm:main_3} in Section~\ref{sec:proofmain3}.

\subsection{The Basics of Strongly Regular Graphs\label{sec:basics}}

Strongly regular graphs are a class of graphs that are generally thought to provide some of the harder instances of graph isomorphism. They arise in various areas of mathematics, and the motivation for studying strongly regular graphs are usually from applications  in experiment design~\citep{bailey}, or due to their geometric and combinatorial aesthetics~\citep{bailey2}.

\begin{defi}[Strongly regular graph]
    A graph is said to be  strongly regular with parameters $(n,d,\lambda,\nu)$ if it has $n$ vertices,  is $d$-regular, and each pair of adjacent (rsp. non adjacent) vertices has exactly $\lambda$ (rsp. $\nu$) common neighbours.
\end{defi}
In general, the parameters $(n,d,\lambda,\nu)$ alone do not determine the isomorphism class of the graph. For example, the Shrikhande graph and the $4\times 4$ rook graph are both strongly regular with parameters $(16,6,2,2)$, but they are not isomorphic. However, strongly regular graphs with the same parameters are cospectral, for the parameters $(n,d,\lambda,\mu)$ uniquely determine the spectrum of the graph\footnote{We refer the reader to Cameron's lecture notes \citep{Cameron2003StronglyRG} for an overview on strongly regular graphs.}.

Latin square graphs are a well-studied class of strongly regular graphs, and their isomorphism problem could very well be a bottleneck for improving the state-of-the-art algorithm for graph isomorphism \citep{babai:babai}, since they naturally arise from finite groups. A \emph{Latin square} $\Lambda$ over an alphabet $A$ is an array of size $|A| \times |A|$,   where each element of the alphabet appears exactly once in  each row and in each column. Formally, we can represent Latin squares as ternary relations $\Lambda\subseteq A^3$, where $(a,b,c)\in \Lambda$ if and only if the $(a,b)$ entry of the array is $c$. The \emph{Latin square graph} associated with the Latin square $\Lambda$ is the graph $G_\Lambda$ whose vertices are the cells of the array (that is, $V\left(G_\Lambda\right)=A^2$) and two such vertices are adjacent if and only if the corresponding cells are in the same row, in the same column, or contain the same element of $A$.

Observe that if $|A|=\alpha$, then $G_\Lambda$ is a graph on $\alpha^2$ vertices; in addition, it can be shown that $G_\Lambda$ is strongly regular with parameters $(\alpha^2,3(\alpha-1),\alpha,6)$. In particular, this implies that Latin square graphs with the same number of vertices are cospectral. A Cayley table of a finite group is an example of a Latin square, though not all Latin squares arise from Cayley tables. There is a natural bijection between isomorphism classes of finite groups and their Latin square graphs, which follows from a well-known theorem \citep{albert1943}.
\begin{thm}[\citep{albert1943}]\label{thm:albert}
    Two finite groups are isomorphic if and only if the Latin square graphs corresponding to their Cayley tables are isomorphic.
\end{thm}
\subsection{Proof of Theorem~\ref{thm:main_3}\label{sec:proofmain3}}
The core of the proof of Theorem~\ref{thm:main_3} is the following combinatorial statement.

\begin{pro}\label{prop:srg}
    Let $G$ and $H$ be strongly regular graphs with parameters $(n,d,\lambda,\nu)$, where $\lambda\geq\nu$. Then, $G$ and $H$ are not isomorphic if and only if  it holds for all $\pi\in\Inj(G,H)$  that $\MMC(\pi)\geq \lambda-\nu+1.$
\end{pro}

\begin{proof}
 If $G$ and $H$ are isomorphic, there must be some $\pi$ for which $\MMC(\pi)=0$. Hence we only need to study the ``only if'' part.
We first assume $\lambda=\nu$. If $G$ and $H$ are not isomorphic, then for any $\pi\in\Inj(G,H)$, there must be an edge $(u,v)\in E(G)$ for which $(\pi(u),\pi(v))\notin E(H)$. As $G$ and $H$ are $d$-regular, by Fact~\ref{fact:pos_neg} the graph $G^\pi-H$ has at least one positive edge and one negative edge incident on the vertex $\pi(u)$, thus giving $\MMC(\pi)\geq 2>1$.

 Now we assume $\lambda>\nu$, and that $(u,v)\notin E(G)$ but $(\pi(u),\pi(v))\in E(H)$. Consider the common neighbourhoods $N_1=N_G(u)\cap N_G(v)$ and $N_2=N_H(\pi(u))\cap N_H(\pi(v))$. From the definition of $\lambda$ and $\nu$, it holds that $|N_1|=\nu$ and $|N_2|=\lambda$, and thus $\pi$ cannot be a bijection between $N_1$ and $N_2$.
In particular, since $\lambda>\nu$, there are at most $\nu$ vertices $w\in N_1$ for which $\pi(w)\in N_2$ and hence, there are at least $\lambda-\nu$ vertices $w'\in V(G)$ such that $\pi\left(w'\right)\in N_2$, but $w'$ is adjacent to at most one vertex between $u$ and $v$. This implies that for such a vertex $w'$, at least one of the edges $\left(\pi\left(w'\right),\pi\left(u\right)\right)$ and $\left(\pi\left(w'\right),\pi\left(v\right)\right)$ is negative for $\pi$. Since $G$ and $H$ are both regular,  by Fact~\ref{fact:pos_neg} for each negative edge incident on some vertex there is a positive edge, and thus a total of at least $2(\lambda-\nu)$ vertices are adjacent to $\pi(u)$ or $\pi(v)$ in $G^\pi-H$. By an averaging argument, it follows that at least one between $\pi(u)$ and $\pi(v)$ has $\lambda-\nu$ incident edges in $G^\pi-H$. These do not take into account the edge $(\pi(u),\pi(v))$ which, by the assumption on $\pi$, is negative in $G^\pi-H$. Hence, at least one vertex between $\pi(u)$ and $\pi(v)$ has at least $\lambda-\nu+1$ incident edges in $G^\pi-H$. Therefore, the statement of the proposition holds.
\end{proof}

 Applying Proposition~\ref{prop:srg} to a Latin square graph with parameters $\lambda=\sqrt{n}$ and $\nu=6$ gives us the following bound, which turns out to be almost tight as shown in Proposition \ref{prop:lsbound} in Section \ref{sec:bound} of the Appendix.

 \begin{cor}\label{cor:ls_mmc}
Let $G$ and $H$ be Latin square graphs on $n$ vertices each. Then, $G$ and $H$ are not isomorphic if and only if it holds  for all $\pi\in\Inj(G,H)$ that $\MMC(\pi)\geq \sqrt{n}-5$.
\end{cor}

Now we are ready to prove Theorem~\ref{thm:main_3}. 

\begin{proof}[of Theorem~\ref{thm:main_3}.]
 Suppose there is a polynomial-time algorithm deciding if $\delta_\edit(G,H)\leq t(n)$ for some $t(n)=o\left(n^{1/2}\right)$ and any pair of $n$-vertex graphs  $G$ and $H$. Since $\mu_1\left(G^\pi,H\right)=\MMC(\pi)$ and $\delta_\edit(G,H)\geq \delta_1(G,H)$, we show that the same algorithm can decide if two Latin square graphs are isomorphic or not based on the following case distinction:
\begin{itemize}
\item If $G$ and $H$ are isomorphic, then $\delta_\edit(G,H)=0\leq t(n)$.
\item If $G$ and $H$ are not isomorphic, then  it holds by Corollary~\ref{cor:ls_mmc} that $\delta_1(G,H)\geq \sqrt{n}-5$. Since $\delta_\edit(G,H)\geq \delta_1(G,H)$ for any graphs $G,H$, it follows that $\delta_\edit(G,H)\geq \sqrt{n}-5$ which is asymptotically larger than $t(n)$.
\end{itemize}
Since the  isomorphism  problem of two  groups as Cayley tables is polynomial-time reducible to the 
isomorphism problem of two Latin square graphs by the proof of Theorem~\ref{thm:albert}, and Latin square graphs with the same number of vertices are cospectral, the first statement of the theorem holds.

The second statement can be obtained in the same way, since the maximum degree of an $n$-vertex Latin square graph satisfies that  $\sqrt{n}=d_{\max}/{3}+1$.
\end{proof}

\section{Further Research Directions\label{sec:open}}
This paper examines  whether well-studied graph metrics can be applied to design approximation algorithms for graph isomorphism, which is a key research problem in theoretical computer science.
While the previous known     \NP-hardness results for the edit and $\ell_p$-operator norms rely on a pair of graphs with different number of edges,  we prove that these \NP-hardness results remain valid even if the two input graphs have the same number of edges. On the other side, it is important to see that our constructed pairs of graphs are still easily distinguishable via efficient means. For example, for our constructed $\hat{G}_k$ and $\hat{H}_k$ from Section~\ref{sec:2ndmain}, the former is bipartite, while the other contains 3-cycles; as such a cospectrality test is sufficient to distinguish the two graphs. This motives the following question:

\begin{openproblem}
Are $\DIST_\edit$, $\DIST_p$ and $\DIST_{|p|}$ \NP-hard on pairs of cospectral graphs?
\end{openproblem}
We partially  address this problem in Section~\ref{sec:srg}, claiming that it is unlikely that $\DIST_\edit$, $\DIST_p$ and $\DIST_{|p|}$ are decidable in polynomial time over pairs of cospectral graphs and, in particular, pairs of Latin square graphs of Cayley tables. In light of Theorem \ref{thm:albert}, it is important to note that, if some similarity problem is \NP-hard on Latin square graphs of Cayley tables, this could translate into an \NP-hard decision problem on finite groups, where these are to be understood as Cayley tables. This would be rather surprising, since all known \NP-hard problems for finite groups usually consider groups presented as a set of generators and their relations, rather than multiplication tables.

A natural follow-up question from Theorem \ref{thm:main_2} is to ask for which functions $t(n)$  is $\delta_\edit(G,H)\leq t(n)$ decidable in polynomial time. Our result addresses such question for polynomials $t(n)$ asymptotically smaller than $\sqrt{n}$. Since any simple graph on $n$ vertices has at most $n(n-1)/2$ edges, it is trivial to show that $\delta_\edit(G,H)\leq n(n-1)/2$ always holds.

\begin{openproblem}
For what values of $c>0$ and $\epsilon>0$ is $\delta_\edit(G,H)\leq cn^\epsilon$ decidable in polynomial time for all pairs of $n$ vertex graphs $G$ and $H$ with equal volumes?
\end{openproblem}

Theorem~\ref{thm:main_2} implies that there are no such values for $\epsilon<1/2$, unless $\textsf{P}=\NP$.

Finally, our discussion in Section~\ref{sec:srg} is centred around a rather niche area of combinatorics and graph theory.  Based on  the construction used in proving Proposition~\ref{prop:lsbound} and the fact that the bound in Corollary~\ref{cor:ls_mmc} might not be tight, we ask the following question.
 
\begin{openproblem}
Given non-isomorphic Latin square graphs $G$ and $H$ on $n$ vertices each, does it always hold that $\delta_1(G,H)\geq \sqrt{n}$?
\end{openproblem}

Given the importance of Latin square graphs in applied areas of mathematics such as experiment design~\citep{bailey} and error correcting codes~\citep{colbourn}, we expect that the answer to the above question could be of interest beyond the pure theory.

%%
%% Bibliography
%%

%% Please use bibtex, 

%\bibliographystyle{acm}
\nocite{*}
%\bibliographystyle{plainnat}
% use the following instead if you encounter problems 

\bibliography{graph_distances}
\label{sec:biblio}

\appendix

\section{Proposition \ref{prop:srg} is tight up to a small constant term}\label{sec:bound}
The combinatorics in the proof of Proposition \ref{prop:srg} is rather crude, so one might expect that a better bound are attainable for certain classes of strongly regular graphs. However, we show that, for  Latin square graphs, the bound is almost tight. Put otherwise, the following statement implies that the bound in Corollary~\ref{cor:ls_mmc} is tight up to a small constant term.

\begin{pro} 
\label{prop:lsbound}
Let $\Gamma$ be a finite group, and consider the non-isomorphic groups $\Gamma\times \mathbb{Z}_4$ and $\Gamma\times\left( \mathbb{Z}_2\right)^2$. Let $G$ and $H$ be the Latin square graphs of their respective Cayley tables. There is an alignment $\pi\in\Inj\left(G,H\right)$ for which $\MMC(\pi)=\sqrt{n}$, where $n=16\left|\Gamma\right|^2$.
\end{pro}

Note that, since Latin square graphs on $n$ vertices are $3\left(\sqrt{n}-1\right)$-regular, Fact~\ref{fact:pos_neg} implies a na\"ive bound of $\MMC(\pi)\leq 6\left(\sqrt{n}-1\right)$, whereas Proposition~\ref{prop:lsbound} implies that the bound in Corollary \ref{cor:ls_mmc} is off by at most a constant term.

To illustrate the idea behind the proof of Proposition \ref{prop:lsbound}, consider the simpler case where $\Gamma$ is the trivial group. First, we introduce some terminology. Consider a Latin square over an alphabet $A$ and recall that the corresponding Latin square graph has vertex set $A^2$. By definition, the vertices $\left(a_1,a_2\right)$ and $\left(b_1,b_2\right)$ are adjacent if one of the following conditions hold:
\begin{enumerate}
    \item $a_1=b_1$, in which case we call the edge between $\left(a_1,a_2\right)$ and $\left(b_1,b_2\right)$ a \emph{row edge}.
    \item $a_2=b_2$, in which case we call to the edge between $\left(a_1,a_2\right)$ and $\left(b_1,b_2\right)$ a \emph{column edge}.
    \item There is some $c\in A$ for which $c$ is the $\left(a_1,a_2\right)$-entry as well as the $\left(b_1,b_2\right)$-entry of the Latin square $\Lambda$. In this case, we call the edge between $\left(a_1,a_2\right)$ and $\left(b_1,b_2\right)$ an \emph{entry edge}.
\end{enumerate}
Note that from the definition of a Latin square, each edge satisfies exactly one of the above conditions.

For simplicity, we consider $\mathbb{Z}_4$ and $\left(\mathbb{Z}_2\right)^2$ as groups over the sets $\left\{0,1,2,3\right\}$ and $\left\{0_s,1_s,2_s,3_s\right\}$ respectively with the same binary operator symbol $+$. Their Cayley tables are reproduced in Table \ref{tab:cayley}.

\begin{table}[h!]
    \vspace{5pt}
    \centering
    \begin{tabular}{c|cccc}
    $+$ & $0$ & $1$ & $2$ & $3$\\
    \hline
    $0$ & $0$ & $1$ & $2$ & $3$\\
    $1$ & $1$ & $2$ & $3$ & $0$\\
    $2$ & $2$ & $3$ & $0$ & $1$\\
    $3$ & $3$ & $0$ & $1$ & $2$\\
\end{tabular}
    \hspace{0.2\textwidth}
    \begin{tabular}{c|cccc}
    $+$ & $0_s$ & $1_s$ & $2_s$ & $3_s$\\
    \hline
    $0_s$ & $0_s$ & $1_s$ & $2_s$ & $3_s$\\
    $1_s$ & $1_s$ & $0_s$ & $3_s$ & $2_s$\\
    $2_s$ & $2_s$ & $3_s$ & $0_s$ & $1_s$\\
    $3_s$ & $3_s$ & $2_s$ & $1_s$ & $0_s$\\
\end{tabular}
    \caption{Cayley tables of $\mathbb{Z}_4$ and $\left(\mathbb{Z}_2\right)^2$.}
    \label{tab:cayley}
\end{table}
If $\Gamma$ is trivial, then $G$ and $H$ are Latin square graphs of the left-most and right-most Cayley tables in Table \ref{tab:cayley} respectively. Thus, $V(G)=\left\{0,1,2,3\right\}^2$ and $V(H)=\left\{0_s,1_s,2_s,3_s\right\}^2$ by the definition of a Latin square graph. We say that the vertices $u=(i,j)\in V(G)$ and $v=\left(i'_s,j'_s\right)\in V(H)$ are \emph{twinned} with one another if $(i+j)_s=i'_s+j'_s$. Put otherwise, $u$ and $v$ are twinned if and only if their corresponding cells in the respective Cayley tables have corresponding entries. For example, $(0,0)$ and $\left(3_s,3_s\right)$ are twinned with one another.

Consider the alignment $\pi\in\Inj\left(G,H\right)$ defined as
$$
(i,j)\mapsto (i_s,j_s)
$$
for all $i,j\in \left\{0,1,2,3\right\}$. It is immediate to see that all the row and column edges of $H$ are neutral for $\pi$, so the only edges in $G^\pi-H$ must be entry edges of $G^\pi$ and $H$. This means that any positive edge $\left(\pi(u),\pi(v)\right)$ in $G^\pi-H$ arises from a pair of edges $u,v\in V(G)$ which are joined by an entry edge; whence, their corresponding cells in the Cayley table have the same entry, but the entries in the cells corresponding to $\pi(u)$ and $\pi(v)$ are different. From the Cayley tables in Table \ref{tab:cayley}, it can be deduced that this only occurs for pairs of vertices $u,v\in V(G)$ whose corresponding cells have even entry and exactly one vertex between $u$ and $v$ is twinned with its image under $\pi$. Given any vertex $u$ with an even number in the corresponding cell, we can construct $4$ such pairs, as predicted by Proposition~\ref{prop:lsbound}. This is because there are exactly $8$ cells with even entry and exactly half of these correspond to vertices that are not twinned with their image under $\pi$.

We now set some notation and terminology for the proof of the Proposition \ref{prop:lsbound}. We identify $\Gamma\times \mathbb{Z}_4$ and $\Gamma\times\left(\mathbb{Z}_2\right)^2$ as groups over the sets $\Gamma\times\left\{0,1,2,3\right\}$ and $\Gamma\times\left\{0_s,1_s,2_s,3_s\right\}$ respectively, and use $\cdot$ to unambiguously denote the operation in both groups. To avoid too many nested brackets, let $g_i$ denote the element $(g,i)\in \Gamma\times\left\{0,1,2,3\right\}$ and similarly, $g_{i_s}$ denotes the element $(g,i_s)\in \Gamma\times\left\{0_s,1_s,2_s,3_s\right\}$. So, for example, $g_i\cdot h_j=(gh)_{i+j}$ where $gh$ is the product in $\Gamma$. We say that the vertices $\left(g_i,h_j\right)$ and $\left(g'_{k_s},h'_{l_s}\right)$ are twinned with one another if $gh=g'h'$ and $\left(i+j\right)_s=k_s+l_s$. We say that the vertex $\left(g_i,h_j\right)\in V(G)$ is even if $i+j$ is even; otherwise, we say it is odd. A similar notion of parity holds for vertices $\left(g_{i_s},h_{j_s}\right)\in V(H)$. Let $G$ and $H$ be the Latin square graphs of Cayley tables of $\Gamma\times\mathbb{Z}_4$ and $\Gamma\times\left( \mathbb{Z}_2\right)^2$ respectively. Thus, $V(G)=\left( \Gamma\times\left\{0,1,2,3\right\}\right)^2$ and $V(H)=\left( \Gamma\times\left\{0_s,1_s,2_s,3_s\right\}\right)^2$. We first prove an auxiliary lemma which generalises our discussion of the case for which $\Gamma$ was taken to be trivial.

\begin{lem}\label{lem:aux}
Let $G$ and $H$ be the Latin square graphs of the Cayley tables of $\Gamma\times \mathbb{Z}_4$ and $\Gamma\times \left(\mathbb{Z}_2\right)^2$. Then,  the following hold:
\begin{enumerate}
    \item The vertices $\left(g_i,h_j\right)$ and $\left(g_{i_s},h_{j_s}\right)$ have the same parity. In particular, all odd vertices are twinned with their image under the mapping $\left(g_i,h_j\right)\mapsto\left(g_{i_s},h_{j_s}\right)$.
    \item An even vertex $\left(g_i,h_j\right)$ is twinned with $\left(g_{i_s},h_{j_s}\right)$ if and only if $i$ and $j$ are both even.
    \item Consider the even vertices $\left(g_i,h_j\right)$ and $\left(g'_k,h'_l\right)$ where $i,j,k,l$ are odd. Then $g_i\cdot h_j=g'_k\cdot h'_l$ if and only if $g_{i_s}\cdot h_{j_s}=g'_{k_s}\cdot h'_{l_s}$.
\end{enumerate}
\end{lem}
\begin{proof}
Note that k the map
$$
\left(g_i,h_j\right)\mapsto\left(g_{i_s},h_{j_s}\right)
$$
preserves the $\Gamma$ component of the vertices, so the three statements above can be verified  directly from the Cayley tables in Table~\ref{tab:cayley}.
\end{proof}

\begin{proof}[of Proposition \ref{prop:lsbound}]
Consider the alignment $\pi\in\Inj(G,H)$ defined as
$$
\left(g_i,h_j\right)\mapsto \left(g_{i_s},h_{j_s}\right)
$$
for all $g,h\in \Gamma,i,j\in\left\{0,1,2,3\right\}$. Note that all the row and column edges of $G^\pi$ and $H$ are neutral for $\pi$, so the only edges appearing in $G^\pi-H$ must be entry edges. From statement $(1)$ of Lemma \ref{lem:aux}, it follows that any odd vertex $\left(g_{i_s},h_{j_s}\right)$ has no incident edges in $G^\pi-H$. Consider vertices $u,v\in V(H)$ where vertices $u=\left(g_{i_s}, h_{j_s}\right)$ and $v=\left(g'_{k_s}, h'_{l_s}\right)$. From statements $(2)$ and $(3)$ of Lemma~\ref{lem:aux} it follows that there is a positive or negative edge in $G^\pi-H$ joining $u$ and $v$ if and only if the following conditions hold:
\begin{itemize}
    \item $gh=g'h'$.
    \item Both $u$ and $v$ are even.
    \item Exactly one of them is twinned with its pre-image under $\pi$.
\end{itemize}
Indeed, if $gh\neq g'h'$, then $(u,v)\notin E(G)$ and $\left(\pi(u),\pi(v)\right)\notin E(H)$. Furthermore, if $u$ and $v$ are both twinned with $\pi(u)$ and $\pi(v)$ respectively, then $\pi$ must preserve their adjacency (that is, whether or not there is an edge between them). Using statement $(3)$ from Lemma \ref{lem:aux}, we deduce that $\pi$ preserves the adjacency between $u$ and $v$ if neither $u$ nor $v$ are twinned with their images under $\pi$.

Consider the case where $u$ is even and twinned with $\pi(u)$. From the above discussion it follows that the number of neighbours of $\pi(u)$ in $G^\pi-H$ is equal to the half the number of cells of the Cayley table of $\Gamma\times \mathbb{Z}_4$ whose entry is equal to $(gh)_0$ or $(gh)_2$. This is because the vertices corresponding to these cells are mapped to a non-twin only if they are in a row labelled by $g''_t$ for some $g''\in \Gamma$ and an odd number $t$. There are $\sqrt{n}/2$ such rows and two such cells for each of such rows, which yields a total of $2\times \sqrt{n}/2=\sqrt{n}$ neighbours of $u$ in $G^\pi-H$ as required. A similar argument shows that if $u$ and $\pi(u)$ are not twinned with one another, then $\pi(u)$ also has exactly $\sqrt{n}$ neighbours in $G^\pi-H$. It then follows that $G^\pi-H$ is $\sqrt{n}$-regular and thus $\MMC(\pi)=\sqrt{n}$ as required.
\end{proof}

\end{document}